\documentclass[12pt,a4paper]{article}
\pdfoutput=1
\usepackage{amssymb}
\usepackage{amsmath}
\usepackage{amsfonts}
\usepackage{epsfig}
\usepackage{xcolor}
\usepackage[hidelinks]{hyperref}
\usepackage{cite}
\usepackage{calligra}
\usepackage[normalem]{ulem}

\usepackage{bm}
\outer\def\beginsection#1\par{\medbreak\bigskip
      \message{#1}\leftline{\bf#1}\nobreak\medskip
\vskip-\parskip
      \noindent}

\numberwithin{equation}{section}

\def\e{\epsilon}

\def\a{\alpha}

\def\d{{\rm d}}

\def\G{\Gamma}

\def\Ord{{\cal O}}

\def\be{\begin{equation}}
\def\ee{\end{equation}}
\def\ba{\begin{eqnarray}}

\def\Imt{{\rm Im}\tau}
\def\Re{{\rm Re}}

\def\G{\mathcal{G}}
\def\V{\mathcal{V}}

\def\mk{\kappa}

\newcommand{\Eq}[1]{Eq.~\eqref{#1}}
\newcommand{\App}[1]{Appendix~\ref{#1}}

\newcommand{\ea}[1]{\begin{align} #1 \end{align}}
\newcommand{\eas}[1]{\begin{align} \begin{split} #1 \end{split} \end{align}}

\newcommand{\seal}[2]{\begin{subequations}\label{#1} \begin{align} #2 \end{align}\end{subequations}}

\usepackage[titles]{tocloft}

\usepackage{titlesec}
\titleformat*{\section}{\large  \bfseries }
\titleformat*{\subsection}{\normalsize  \bfseries }

\begin{document}

\begin{titlepage}
\hfill \hbox{NORDITA-2018-068}
\vskip 1.5cm
\begin{center}
{\Large \bf
Multiloop Soft Theorem for Gravitons and Dilatons
in the Bosonic String
}

\vskip 1.0cm {\large Paolo
Di Vecchia$^{a,b}$,
Raffaele Marotta$^{c}$, Matin Mojaza$^{d}$
} \\[0.7cm] 
{\it $^a$ The Niels Bohr Institute, University of Copenhagen, Blegdamsvej 17, \\
DK-2100 Copenhagen \O , Denmark}\\
{\it $^b$ Nordita, KTH Royal Institute of Technology and Stockholm 
University, \\Roslagstullsbacken 23, SE-10691 Stockholm, Sweden}\\[2mm]
{\it $^c$  Istituto Nazionale di Fisica Nucleare, Sezione di Napoli, Complesso \\
 Universitario di Monte S. Angelo ed. 6, via Cintia, 80126, Napoli, Italy}\\[2mm]
 {\it $^d$  Max-Planck-Institut f\"ur Gravitationsphysik, \\
Albert-Einstein-Institut, Am M\"uhlenberg 1, 14476 Potsdam, Germany}
\end{center}
\begin{abstract}
We construct, in the closed bosonic string,  the multiloop amplitude involving $N$ tachyons and one massless particle with $26 -D$ compactified directions,
and we show that at least for $D>4$, the soft behaviors of the graviton and dilaton satisfy the same soft theorems as at the tree level, up to one additional term at the subsubleading order, which can only contribute to the dilaton soft behavior and which we show is zero at least at one loop. This is possible, since
the infrared divergences due to the non-vanishing tachyon and dilaton tadpoles do not depend on the number of external particles and are therefore the same both in the amplitude with the soft particle and in the amplitude without the soft particle.
Therefore this leaves unchanged the soft operator acting on the amplitude without the soft particle.
The additional infrared divergence 
appearing  for $D \leq 4$ depend on the number of external legs and must be understood on their own.

\end{abstract}

{\let\thefootnote\relax
\footnotetext{divecchi@nbi.ku.dk, raffaele.marotta@na.infn.it, matin.mojaza@aei.mpg.de}
}

\vfill

\end{titlepage}


\setlength{\parskip}{5mm plus2mm minus2mm}

\tableofcontents

\vspace{5mm}
\clearpage

\setlength{\parskip}{3mm plus2mm minus2mm}

\section{Introduction}
\label{intro}
Soft theorems for gravitons and other massless particles at the tree level have been intensively studied in the last few years both in field theory~\cite{field} and in string theory~\cite{string}. In particular, in string theory, it has been shown that,
			 the leading,  subleading and subsubleading behavior of a soft dilaton is universal, i.e. it is the same  in any string theory, while, in the case of the graviton, the subsubleading behavior has, in general, string corrections that    depend on the string theory under consideration\cite{DiVecchia:2015jaq,1610.03481}. In both cases those soft theorems are a direct consequence of gauge invariance\cite{gaugeinvariance} and of the structure of the three-point vertex containing a soft particle and two hard particles\cite{1604.03355}. Gauge invariance fixes also the leading 
 soft behavior  of the  Kalb-Ramond field~\cite{1706.02961}.

At  loop level it has been observed that the tree soft theorems are modified by infrared and ultraviolet divergences occurring in field theory\cite{Bern:2014oka,He:2014bga,Broedel:2014bza}, but, if one considers   gravity theories that are ultraviolet complete and free from infrared divergences, by taking  the number of space-time dimensions $D>4$, then one gets again a universal soft behavior for gravitons up to subleading level~\cite{Sen:2017nim},
and a factorizing soft behavior through subsubleading order, where universality is broken only by the two- and three-point coupling of the soft graviton to the other fields of the underlying theory~\cite{Laddha:2017ygw}.

In this paper we consider the $h$-loop  amplitude of the closed bosonic string involving  $N$ tachyons and one massless state and we show that we obtain the same soft theorems for gravitons, 
as at tree-level, as well as for the dilaton once its soft operator is rewritten in terms of the scaling properties of the amplitude, as long as we keep the non-compact number of the space-time dimensions, $D$, to be greater than four,
and up to  possibly  an additional term at the subsubleading order, which, however,  we have been able to show  to be zero  at one loop.

As in Refs.~\cite{Sen:2017nim,Laddha:2017ygw} we need  $D>4$ in order to avoid infrared divergences that depend on the number of external legs~\cite{Bern:2014oka}. On the other hand, working in a string theory, we have no ultraviolet divergences. We have, however,  in the bosonic string infrared divergences due to the dilaton tadpole. These divergences are, however, not dependent on the number of the external states and therefore  appear both in the multiloop amplitude with $N$ tachyons and one massless state and in the amplitude with only $N$ tachyons leaving the soft operator connecting them 
{unaffected by these divergences.}

The $h$-loop amplitude with $N$ tachyons and one massless state is obtained from the $h$-loop $(N+1)$-Reggeon vertex constructed in Refs.~\cite{DiVecchia:1987ew,DiVecchia:1988cy,DiVecchia:1988hq},  by means of the sewing procedure, starting from the tree diagram $(N+1)$-Reggeon vertex originally constructed by 
Lovelace~\cite{Lovelace:1970ej} including the part with the ghost coordinates~\cite{DiVecchia:1986uu,DiVecchia:1987ew}.
Multiloop amplitudes in the bosonic string were computed even before one realized that the extension of the Veneziano model was a string theory. The correct integration measure over the moduli was, however, only fixed  in the eighties after the formulation of the BRST invariant action for the bosonic string theory~\cite{DiVecchia:1987uf,Petersen:1987pv,Mandelstam:1985ww,Mandelstam:1991tw}.  It turns out that the $h$-loop amplitude with $N$ tachyons and one massless state, $\mathcal{M}_{N+1}^{(h)}$, except for one term, which is present only in the dilaton amplitude and  that we will discuss later, has the same form in terms of the Green function, as in the case of the tree diagrams. This observation allows us to obtain the graviton soft theorem at the multiloop level as it was done in Ref.~\cite{DiVecchia:2015oba} at the tree level. Because of this, the soft operator is the same at tree and loop level.

 In the case of a soft  dilaton  the presence of the extra term 
 gives a new type of contribution starting from the subleading order.
 We have been able to evaluate it at subleading order at the multiloop level,
 leading to the following soft theorem up to subleading order:
\ea{
M_{N;\phi}^{(h)} (k_i; q) = 
\frac{\kappa_D}{ \sqrt{D-2} }  \left[  - \sum_{i=1}^{N} \frac{ m^2}{k_i q}{e^{q\partial_{k_i}}} + 2 - \sum_{i=1}^N k_i \cdot \partial_{k_i}
+ h  (D-2)
   \right]  M_N^{(h)} + \Ord(q) \, , 
}
where $\kappa_D$ is the gravitational constant in $D$ non-compact space-time dimensions {(see App.~\ref{App:kappaD} for a definition)},
$m^2 = - \frac{4}{\alpha'}$ is the mass of the tachyons, and the operator ${e^{q\partial_{k_i}}}$ should be considered expanded up to $\Ord(q)$. Apart from the $h$-dependent term, this is the same operator as at tree-level.
 
At subsubleading order, we have evaluated the contribution from the extra term only at one loop, where it turns out to vanish as a consequence of momentum conservation. Hence the tree-level soft theorem is at this order unchanged at the one loop level, where the soft theorem reads
\ea{
M_{N;\phi}^{(1)} (k_i; q) = 
\frac{\kappa_D}{ \sqrt{D-2} }  \left[ - \sum_{i=1}^N \frac{m^2}{k_i q} {\rm e}^{q \partial_{k_i}}  + D - \sum_{i=1}^N\hat{D}_i
+  q_\mu \sum_{i=1}^N {\hat{K}}_{i}^{ \mu}
   \right]  M_N^{(1)}    + \Ord(q^2) \, , 
}
where we defined
\ea{
\hat{D}_i = k_{i} \cdot  
\frac{\partial}{\partial k_{i}}  \, , \qquad 
 \hat{K}_{i}^{\mu} = \frac{1}{2} k_{i}^{ \mu} \frac{\partial^2}{\partial
k_{i\nu} \partial k_i^\nu} 
-k_{i}^{\rho} \frac{\partial^2}{ \partial k_{i}^{\rho} \partial
k_{i\mu}}  \, ,
}
the generators of the space-time dilatations and special conformal transformations. 

The subleading soft operator differs 
from the tree-level one by the term explicitly dependent on the loop number $h$, which at first sight looks like an obstruction against a soft theorem for the full amplitude.
It is, however, possible to recast the $h$-loop soft dilaton operator 
into a form that makes it the same at any number of loops. 
To this end, recall that
the string amplitudes depend on three constants; the 
Regge  slope $\alpha'$, the string coupling constant $g_s$ and the gravitational coupling constant $\kappa_D$.
 Only two of them are fundamental and one can take $\kappa_D$ to be a function of the other two constants. 
{In a compactified theory, there is additionally a dependence on the compactification parameters, such as the radii of the compact manifold, which we collectively denote by $R$, and $\kappa_D$ can be considered also a function of those parameters (see App.~\ref{App:kappaD}).}
By taking into account the explicit dependence of the scattering amplitudes on these fundamental constants, we can rewrite the dilatation operator entering the $h$-loop soft dilaton operator as a scaling operator in terms of the fundamental constants, as also originally done at tree-level in Ref.~\cite{Ademollo:1975pf}. In this way, one exactly gets rid of the $h$-dependence, and finds the following dilaton soft theorem, {valid to subleading order,} for the full all-loop amplitude, $\mathcal{M}_N = \sum_{h=0}^\infty M_N^{(h)}$, when the other particles are $N$ closed string tachyons:
\ea{
\mathcal{M}_{N;\phi} (k_i; q) = &
\frac{\kappa_D}{ \sqrt{D-2} } 
\left[ - \sum_{i=1}^{N} \frac{ m^2}{k_i q}e^{q\partial_{k_i}} +\frac{D-2}{2} g_{s}\frac{\partial}{\partial g_{s}} - \sqrt{\alpha'} \frac{\partial}{\partial \sqrt{\alpha'}}  {- R \frac{\partial }{\partial R} }
\right] \mathcal{M}_N(k_i)
+ \Ord(q)\, .
}

The paper is organized as follows. In Sect. \ref{NReggeon} we discuss the $h$-loop $N$-Reggeon vertex and from it we derive the $h$-loop amplitude involving $N$ tachyons and one massless state.
In Sect. \ref{Soft} we derive the soft behavior of the amplitude. {In Sect. \ref{IRD} we  discuss   the infrared divergences both those due to the tadpoles of the bosonic string amplitudes and the ones appearing when the theory is compactified to four dimensions. In Sect. \ref{conclusion} we present our conclusions. 
{In Appendix~\ref{App:kappaD} a derivation of the gravitational coupling constant in the compactified theory and its relation to $g_s$ and $\alpha'$ is given.} In Appendix~\ref{Schottky} we review the Schottky parametrization of Riemann surfaces, and in Appendix~\ref{AppB} we discuss some properties of the multiloop Green function.
Appendix \ref{NRegge}, \ref{CalS1}, and \ref{X}, give calculational and technical details on expression given in the text.

\vspace{-3mm}

\section{Multiloop amplitude in the bosonic string}
\label{NReggeon}
\setcounter{equation}{0}
\vspace{-4mm}

In this section we compute in the bosonic string,  toroidally compactified on  $\mathbb{R}^{1,D-1}\otimes \mathbb{T}^{26-D}$,   the multiloop amplitude  containing $N$ tachyons and one massless state from the multiloop $N$-Reggeon vertex. In particular, in the first subsection we
write down the $N$-Reggeon vertex. It is derived from previous literature~\cite{DiVecchia:1988cy,DiVecchia:1987uf}  and  some details of its derivation are put in  Appendix \ref{NRegge}. In  the second subsection we use it to construct the multiloop amplitude involving $N$ tachyons and one massless state {in the closed bosonic string}. 

\subsection{The $h$-loop  $N$-Reggeon vertex}
\label{NReggeon1}

We start by writing  the $N$-Reggeon vertex for the  closed bosonic string~\cite{DiVecchia:1988cy}  (see \App{NRegge} for details of the derivation from the expression in Ref.~\cite{DiVecchia:1988cy}):
\begin{eqnarray}
\V_N = && C_h (N_0)^N
\int dV_N \langle \Omega | 
\exp \left[\frac{1}{2} \sum_{i=1}^N \sum_{n=0}^\infty \frac{\alpha_n^{(i)}}{n!} \alpha_0^{(i)} \frac{\partial^n}{\partial z^n} \log V_i ' (z) \Big|_{z=0}\right]
\nonumber \\
&&  \times
\exp \left[\frac{1}{2} \sum_{i=1}^N \sum_{n=0}^\infty \frac{{\bar{\alpha}}_n^{(i)}}{n!} \alpha_0^{(i)} \frac{\partial^n}{\partial {\bar{z}}^n} \log {\bar{V}}_i ' ({\bar{z}}) \Big|_{{\bar{z}}=0}\right]
\nonumber \\
&& \times \exp \left[  \frac{1}{2}\sum_{i \neq j} \sum_{n,m=0}^{\infty} \frac{\alpha_{n}^{(i)}}{n!}  \partial_z^n \partial_y^m \log \frac{E ( V_i (z) , V_j (y) )}{\sqrt{V_i ' (0) V_j ' (0)}}\Big|_{z=y=0} \frac{\alpha_m^{(j)}}{m!} \right] \nonumber \\
&& \times \exp \left[  \frac{1}{2}\sum_{i \neq j} \sum_{n,m=0}^{\infty} \frac{{\bar{\alpha}}_{n}^{(i)}}{n!}  \partial_{\bar{z}}^n \partial_{\bar{y}}^m \log \frac{E ( {\bar{V}}_i ({\bar{z}}) , {\bar{V}}_j ({\bar{y}}))}{
\sqrt{{\bar{V}}_i' (0) {\bar{V}}_j ' (0)}} \Big|_{z=y=0} \frac{{\bar{\alpha}}_m^{(j)}}{m!} \right] \nonumber \\
&& \times \exp \left[ \frac{1}{2} \sum_{i=1}^N \sum_{n,m=0}^{\infty} \frac{\alpha_n^{(i)}}{n!} \partial_z^n \partial_y^m \log \frac{ E (V_i (z) , V_{i} (y) )}{V_i (z) - V_i (y) }\Big |_{z=y=0}  \frac{\alpha_m^{(i)}}{m!}\right]
\nonumber \\
&& \times \exp \left[ \frac{1}{2} \sum_{i=1}^N
\sum_{n,m=0}^{\infty} \frac{{\bar{\alpha}}_n^{(i)}}{n!}  \partial_{\bar{z}}^n \partial_{\bar{y}}^m \log \frac{ E ({\bar{V}}_i ({\bar{z}}) , {\bar{V}}_{i} ({\bar{y}}) )}{{\bar{V}}_i ({\bar{z}}) - {\bar{V}}_i ({\bar{y}}) }\Big |_{{\bar{z}}={\bar{y}}=0}  \frac{{\bar{\alpha}}_m^{(i)}}{m!}\right]
\nonumber \\
&& \times \exp \left[ \sum_{i, j=1}^{N} \sum_{n=0}^{\infty} \left( \frac{\alpha_n^{(i)}}{n!} \partial_z^n +
\frac{{\bar{\alpha}}_n^{(i)}}{n!} \partial^n_{ {\bar{z}} } \right) \Re \left( \int_{z_0}^{V_i (z)} \omega_I 
\right)  (2\pi \Imt)^{-1}_{IJ}  \right. \nonumber \\
 && \qquad \qquad \times \left.  \sum_{m=0}^{\infty} \left( \frac{\alpha_m^{(j)}}{n!} \partial_z^m +
\frac{{\bar{\alpha}}_m^{(i)}}{m!} \partial^m_{ {\bar{z}} } \right) \Re \left( \int_{z_0}^{V_j (y)} \omega_J 
\right)  \right] \, ,
\label{NL1}
\end{eqnarray}
where 
\begin{eqnarray}
\alpha_n = a_n \sqrt{n} \,\ \ \text{if} \,\ \ n \neq 0~,~~~ \alpha_0 = {\bar{\alpha}}_0 = \frac{\sqrt{2 \alpha'}}{2} p \, , 
\label{NL2}
\end{eqnarray}
with $p$ being at this level still an operator. The functions $V_i (z)$, satisfying the condition $V_i (0) = z_i$, parametrize the coordinates around the various punctures (see Appendix \ref{CooChoice} for details). 
It can be seen that, if the external states are on-shell physical states, the dependence on the $V_i (z)$ drops out. In the following  we keep, however, the variables $V_i (z)$ in order  to define a proper Green function.
The quantity $E(z,y)$ is the \emph{prime form}, $\omega_I$ are  the \emph{abelian differentials} for \mbox{$I=1 \dots h$} with $h$ being the genus  of the Riemann surface, and $\tau$ is the \emph{period matrix}. Repeated capital indices $I, J, \ldots$ are assumed to be summed over from $1, \ldots, h$. All these quantities  are defined and discussed in \App{Schottky}.
The constants $C_h$ and $N_0$ provide the correct normalization of the amplitude, and are equal to
\ea{
C_h = C_0\,N_0^{2h}\, \left(\frac{\alpha'}{8\pi}\right)^h \frac{1}{ (2\pi \alpha' )^{\frac{D h}{2}}}
=
\left( \frac{8\pi}{\alpha'}\right)^{1-h} \left( \frac{\kappa_{D}}{2\pi}\right)^{2(h-1)} 
\frac{1}{ (2\pi \alpha' )^{ \frac{hD}{2}}} 
\quad ; \quad
N_0 = \frac{\kappa_D}{2\pi} \, .
\label{NL2c}
}
These expressions follow
 from the  sewing procedure, which allows to obtain  $h$-loop amplitudes from tree-level ones by sewing  together $2h$ external legs  with the propagator 
\begin{eqnarray}
\label{propa}
\frac{\alpha'}{8\pi} \int_{|z|\leq 1} \frac{d^2 z}{|z|^2} z^{L_0-1}\bar{z}^{\bar{L}_0-1} \, ,
\end{eqnarray}
to produce $h$ handles (or loops). 
The first factor, $C_0$, on the left hand side of  Eq.~(\ref{NL2c}) takes in account the normalization  
of the tree-level amplitude, the second factor, $N_0^{2h}$, is the normalization of the sewed $2h$ legs, the third  factor, $(\alpha'/8\pi)^h$, comes from  the normalization of the propagator in Eq.~(\ref{propa}), while the last factor, $(2\pi \alpha')^{-D h/2}$, arises from the integration over  the momenta circulating in the loops.

The measure of the moduli, in the Schottky parametrization of the Riemann surface, including the compactification factor, is equal to~\cite{DiVecchia:1987uf,Petersen:1987pv,Mandelstam:1985ww,Mandelstam:1991tw}
\footnote{In this work we use the convention $d^2 z_i=2\d\Re(z_i) \d{\rm Im}(z_i)$.
}  
\ea{
d V_N = &\prod_{i=1}^{N} \left(\frac{ d^2 z_i}{|V_i ' (0)|^2}\right) \frac{1}{dV_{abc}} \prod_{I=1}^h \left[ \frac{ d^2 \mk_I d^2 \xi_I d^2 \eta_I}{|\mk_I|^4|\xi_I - \eta_I |^4} |1- \mk_I |^4 \right] \left( \det 2\pi   \Imt \right)^{-\frac{D}{2}}
\nonumber \\
&\times \prod_{\alpha} ' \left[ \prod_{n=1}^{\infty} \left | \frac{1}{1 -\mk_\alpha^n} \right |^{52} \prod_{n=2}^{\infty} | 1 - \mk_\alpha^n |^4  \right] 
[F(\tau,\,\bar{\tau})]^{26-D}
\, .
\label{NL2a}
}
Here $d V_{abc}$ is the volume element of the $SL(2, \mathbb{C})$ M\"obius group, $\xi_I$ and $\eta_I$ are the attractive and repulsive fixed points  and  $\mk_I$ is the multiplier
of the $h$ generators of the Schottky group%
\footnote{We hope that the multipliers $\mk_I$ are not confused with the gravitational constant $\kappa_D$.}. 
$\mk_\alpha$  is the multiplier of a primary class and $\prod_\alpha '$ is a product over primary classes (see Appendix E of Ref.~\cite{DiVecchia:1988cy} for details).
The factor $\left( \det (2\pi \Imt) \right)^{-\frac{D}{2}}$ in the measure  comes  from the integral over the momenta along the non-compact directions circulating in the  loops. For the momenta along the compact dimensions one must replace  the integral over the momenta  with  a sum over the Kaluza-Klein modes $\bf{n}$ and  the winding numbers $\bf{m}$. Therefore, for each compact dimension, the factor $\left( \det (2\pi \Imt) \right)^{-\frac{1}{2}}$ is replaced  by \cite{Giveon:9401139}:
\begin{eqnarray}
F(\tau,\,\bar{\tau})=
\sum_{({\bf m},{\bf n})\in \mathbb{Z}^{2h}}e^{i\pi\left({\bf p_R}\tau{\bf p_R}-{\bf p_L}\bar{\tau}{\bf p_L}\right) }
\, ,
\label{compact}
\end{eqnarray}
where ${\bf p_{R;L}}=\frac{1}{\sqrt{2}} \left(\frac{\sqrt{\alpha'}}{R} {\bf n} \pm\frac{R}{\sqrt{\alpha'}}{\bf m}\right)$. 
{Here $R$ denotes collectively the compactification radii.}

Finally, the vacuum state is defined by:
\ea{
\langle \Omega | \equiv \prod_{i=1}^{N} [ {}_i \langle x=0; 0_a , 0_{\bar{a}} |] (2\pi)^{D} \delta^{(D)} 
\left( \textstyle \sum_{i=1}^n p_i \right) \, .
\label{NL2b}}
It should in principle  also depend on the winding numbers and Kaluza-Klein modes of the compact dimension, but they are now irrelevant, and thus suppressed, since, in our case, the external states have momenta and oscillators only along non-compact directions.

We conclude this section by observing that in a compact space, the $h$-loop $N$-Reggeon Vertex depends also on the annihilation operators associated to the $26-D$ compact directions and on the left and right compact momentum operators. These should be included  in Eq.~(\ref{NL1}) but we have  neglected them because they are  irrelevant in our calculation. The physical states, tachyons and massless states of the closed string, considered in this paper are vacuum states along the compact directions with zero winding number  and Kaluza-Klein momenta. Therefore  there are no contributions to the amplitude coming from these compact degrees of freedom.
The only dependence from the compact directions is the one due to  right and left discrete momenta circulating in the loop which has been properly taken in account in Eq.~(\ref{compact}).

\subsection{$h$-loop amplitude with $N$ tachyons and one massless state}
\label{ntachyon}

We now specialize the vertex in \Eq{NL1} for tachyons and massless states.
We make the transition $N \to N+1$, since we will in the end specify $N$ states to be tachyons and one to be a massless state, but for now the $N+1$ states can be any of the two.
For these states the vertex in \Eq{NL1} reduces to
\begin{eqnarray}
&&  \V_{N+1}=  C_h (N_0)^{N+1}
\int dV_{N+1} \langle \Omega| 
\exp \left[\frac{1}{2} \sum_{i \neq j=1}^{N+1} \left (  \frac{\sqrt{2\alpha'}}{2} p_i + V_i ' (0) a_{1}^{(i)} \partial_{z_i} 
+{\bar{V}}_i ' (0) {\bar{a}}_{1}^{(i)} \partial_{{\bar{z}}_i} \right  ) \right. \nonumber \\
&& \times \left.
\left   (  \frac{\sqrt{2\alpha'}}{2} p_j + V_j ' (0) a_{1}^{(i)} \partial_{z_j} +
   {\bar{V}}_j ' (0) {\bar{a}}_{1}^{(j)} \partial_{{\bar{z}}_j}\right ) 
   \log  \frac{\left |E (z_i , z_j )\right |^2}{|V_i ' (0) V_j ' (0)|}  \right]
\nonumber \\
&&
\times \exp \Bigg[ \sum_{i,j=1}^{N+1} \left( \frac{\sqrt{2\alpha'}}{2} p_i + V_i ' (0) a_1^{(i)} \partial_{z_i} + 
{\bar{V}}_i ' (0) {\bar{a}}_1^{(i)} \partial_{{\bar{z}}_i}  \right) \Re \left( \int_{ z_0}^{z_i} \omega _I \right) 
( 2\pi \Imt)^{-1}_{IJ} \nonumber \\
&& \times
  \left( \frac{\sqrt{2\alpha'}}{2} p_j + V_j ' (0) a_1^{(j)} \partial_{z_j} + 
{\bar{V}}_j ' (0) {\bar{a}}_1^{(j)} \partial_{{\bar{z}}_j}  \right)  \Re \left( \int_{ z_0}^{ z_j} \omega _J\right) 
\Bigg]   \, .
\label{NL3}
\end{eqnarray}
The exponentials in the first, second, fifth and sixth lines of  Eq.~(\ref{NL1}) do not contribute
for the tachyon states. They also do not contribute  for the massless states because in this case one obtains terms proportional to $(q \epsilon)$ that are  zero for physical massless states ($q$ is the momentum of the massless state and $\epsilon$ its polarization). 
 
Separating in the last two lines terms with $i \neq j$ from those with $i=j$  and eliminating the dependence on $z_0$ as showed in \App{NRegge},  we can write the previous equation as follows:
\ea{
 \V_{N+1}=  C_h (N_0)^{N+1} \int dV_{N+1} &\langle \Omega| 
 \exp \left[ \frac{1}{2} \sum_{i \neq j=1}^{N+1}  \left( \frac{\sqrt{2\alpha'}}{2} p_i + V_i ' (0) a_1^{(i)} \partial_{z_i} + 
{\bar{V}}_i ' (0) {\bar{a}}_1^{(i)} \partial_{{\bar{z}}_i}  \right)  \right. 
\nonumber \\
& \times \left. \left( \frac{\sqrt{2\alpha'}}{2} p_j+ V_j ' (0) a_1^{(j)} \partial_{z_j} + 
{\bar{V}}_j ' (0) {\bar{a}}_1^{(j)} \partial_{{\bar{z}}_j}  \right)  {\cal{G}}_h (z_i , z_j )\right]
\nonumber \\
& \times \exp \left[ \frac{1}{2} \sum_{i=1}^{N+1} |V_i ' (0)|^2 \omega_I (z_i )  (2\pi \Imt )^{-1}_{IJ}
{\bar{\omega}} ({\bar{z}}_J ) \alpha_1^{(i)} {\bar{\alpha}}_1^{(i)} \right] \, ,
\label{NL5}
}
where
\begin{eqnarray}
{\cal{G}}_h (z_i , z_j ) = \log \frac{  | E(z_i , z_j )|^2}{|V_i ' (0) V_j ' (0)|}  + \Re \left( \int_{z_j}^{z_i} \omega _I\right) 
( 2\pi \Imt)^{-1}_{IJ} \Re \left( \int_{z_i}^{z_j} \omega _J \right) \, .
\label{NL6}
\end{eqnarray}

When evaluated on on-shell external states the final result for the amplitude will, as a consequence of momentum conservation, not depend on the factor $|V_i (0) V_j ' (0)|$ included in $\G_h$.
We have, however, included them in the definition of $\G_h$, since this allows us to identify it with the regularized Green function of Ref.~\cite{DHoker:1988pdl}.
As shown in \App{CooChoice},
by choosing conformal coordinates with metric
$ds^2 = \rho(z, \bar{z}) dz d\bar{z}$ to parametrize the Riemann surface around the punctures $z_i$, the functions $|V_i'(0)|^2$ can be set equal to
\ea{
|V_i'(0)|^2 = \frac{1}{\rho (z_i, \bar{z}_i)} \, .
\label{coordinatechoice}
}
With this choice, the function $\G_h$ is exactly equal to minus the regularized Green function $G_r$ discussed in \App{AppB}, from where it follows that $\G_h$ then satisfies
\ea{
\partial_z \partial_{\bar{z}} \mathcal{G}_h(z, w)  = \pi \delta^{(2)}(z-w) - &\frac{1}{2} \omega_I (z) (2\pi {\rm Im} \tau )_{IJ}^{-1} \, \bar{\omega}_J(\bar{z})
+ \frac{1}{2}\partial_z \partial_{\bar{z}} \log \rho(z, \bar{z}) \, ,
\label{d2G}\\[2mm]
&\int \d^2 z \partial_z \partial_{\bar{z}} \mathcal{G}_h(z, w)  = 0 \, .
\label{Intd2G}
}
In the rest of this paper we shall assume 
the choice of coordinates  in Eq.~(\ref{coordinatechoice})}.

The previous vertex is valid for any number of tachyons and massless states. In the following we restrict ourselves  to the case of $N$ tachyons and one massless state. This means that we have to saturate it with the states given by~\footnote{%
We are again suppressing the vacuum structure along the compact directions,
but as discussed at the end of sect.~\ref{NReggeon1}, for our purposes they are irrelevant.
} 
\begin{eqnarray}
\prod_{i=1}^{N}  [ | 0 , k_i \rangle ] a_{1 \mu }^{\dagger } {\bar{a}}_{1 \nu}^{\dagger} | 0, q \rangle \, ,
\label{NL3a}
\end{eqnarray}
where the   $N$ tachyons have momenta $k_i$ and the massless state momentum $q$.

After careful contractions and some rewriting we get the following expression for the  $h$-loop amplitude for $h \geq 1$:
\ea{
 M_{N; 1}^{(h)} =& C_h (N_0)^{N+1} \int dV_N  \prod_{i<j=1}^N  {\rm e}^{ \frac{\alpha'}{2} k_i k_j  \G_h (z_i , z_j ) }\epsilon^\mu_q  {\bar{\epsilon}}^\nu_q \int d^2 z \prod_{\ell=1}^N
{\rm e}^{ \frac{\alpha'}{2}  k_{\ell} q \G_h (z, z_{\ell})} 
  \nonumber \\
& \times \left[ \frac{\alpha'}{2} \sum_{i,j=1}^N k_{i\mu} k_{j\nu}   \partial_z \G_h (z, z_i ) 
\partial_{\bar{z}} \G_h (z, z_j ) + \frac{1}{2} \eta_{\mu \nu} \omega_I (z) ( 2\pi \Imt )^{-1}_{IJ}
{\bar{\omega}}_J (z)  \right]\, .
\label{ML3a}
}

{In comparison, the tree-amplitude for the scattering of a massless particle and $N$ tachyons in the bosonic string is given by
\begin{eqnarray}
M_{N;1}^{(0)} = &&C_0 (N_0)^{N+1} \int \frac{\prod_{i=1}^N d^2 z_i}{dV_{abc}} 
 \prod_{i<j=1}^N {\rm e}^{ \frac{\alpha'}{2} k_i k_j  \G_0 (z_i , z_j ) } 
\epsilon^\mu_q {\bar{\epsilon}}^\nu_q  \int d^2 z \prod_{\ell=1}^N 
{\rm e}^{ \frac{\alpha'}{2}  k_i q \G_0 (z, z_{\ell})} 
 \nonumber \\
&& \times  \frac{\alpha'}{2}\sum_{i,j=1}^{N}  k_{i\mu}   k_{j\nu}  
\partial_z \G_0 (z, z_i ) \partial_{\bar{z}} \G_0 (z, z_j )\, ,
\label{ML1}
\end{eqnarray}
where $\G_0 (z,w) = \log |z-w|^2$.
Except for the second term in the square bracket in Eq.~(\ref{ML3a}), and for the integration measure and the integration region,
the two expressions in Eqs.~(\ref{ML1}) and (\ref{ML3a}) have the same form in terms of their Green function. Notice that the extra term contributes only when the massless state is a  dilaton.


\section{Soft behavior of a massless closed string at multiloops}
\label{Soft}
\setcounter{equation}{0}

Starting from \Eq{ML3a}, we would like to study its soft behavior when the momentum carried by the massless state  is much lower than the momenta of the tachyons. Let us first notice that the $h$-loop amplitude can be separated into a part describing just the tachyon scattering convoluted with the contributions to the scattering of the massless state
i.e.
\seal{ConvolutedAmplitude}{
M_{N;1}^{(h)} &=  M_N^{(h)} \ast S (q, k_i; z_i ) \, ,\\[2mm]
M_N^{(h)} &= 
C_h N_0^{N} \int
dV_N
\prod_{i< j}^N e^{ \frac{\a'}{2} k_i k_j  \mathcal{G}_h(z_i, z_j) } \, ,
\label{3.1b}\\
S(q, k_i; z_i ) &=
N_0 \int \d^2 z  \, 
\prod_{i=1}^N e^{ \frac{\a'}{2} k_i q  \mathcal{G}_h(z_i, z) }
\nonumber \\[2mm]
& \quad \times
\int \d^2 \theta
\exp \left \{ \sum_{i=1}^{N} \sqrt{\frac{\a'}{2}}k_i
\left (\theta \e \partial_{z} + \bar{\theta } \bar{\e} \partial_{\bar{z}} \right )
 \mathcal{G}_h(z_i, z) 
\right \}
\nonumber \\
&\quad \times
\exp \left \{ \frac{1}{2} \theta \e \cdot \bar{\theta } \bar{\e} 
\  \omega_I(z) (2 \pi \, \Imt )_{IJ}^{-1} \, \bar{\omega}_J(\bar{z})
 \right \} \, ,
 }
where $\ast$ denotes a convolution of the integrals. 
On its own, $M_N^{(h)}$ is exactly the $h$-loop $N$-tachyon amplitude. The Grassmanian integral in $S$ is easy to perform, yielding
\ea{
S(q,k_i; z_i) = N_0
\int \d^2 z  \, 
\prod_{i=1}^N e^{ \frac{\a'}{2} k_i q  \mathcal{G}_h(z_i, z) }
&\left [
\frac{\a'}{2} \sum_{i,j=1}^N (k_i \e) (k_j \bar{\e}) \partial_{z} \mathcal{G}_h(z_i,z)\partial_{\bar{z}} \mathcal{G}_h(z_j,z)
\right .
\nonumber \\
&\left . 
+
 \frac{1}{2}  (\e \cdot \bar{\e}) 
 \ \omega_I(z) (2 \pi \, \Imt )_{IJ}^{-1} \, \bar{\omega}_J(\bar{z})
\right ]\, .
}
We would like to compute $S$ through order $q$ in the soft momentum limit.
The second term above vanishes for the graviton and the Kalb-Ramond antisymmetric field, but contributes in the case of the dilaton (because the polarization tensor is traced). 
We separate the two terms accordingly into $(N_0 S_1)$ and $(N_0 S_2)$, and expand them (partly) in $q$ as follows
\ea{
S_1 = 
\int \d^2 z  \, 
&\left [ 1 + \sum_{j\neq i}^N  \frac{\a'}{2} k_j q  \mathcal{G}_h(z_i, z) + \frac{1}{2} \left (\frac{\a'}{2}\right )^2 
\sum_{j, l\neq i}^N  (k_j q) (k_l q)  \mathcal{G}_h(z_j, z) \mathcal{G}_h(z_l, z) \right ]
\nonumber \\
&\times
\frac{\a'}{2} \sum_{i,j=1}^N   (k_i \e) (k_j \bar{\e}) \partial_{z} \mathcal{G}_h(z_i,z)\partial_{\bar{z}} \mathcal{G}_h(z_j,z)e^{ \frac{\a'}{2} k_i q  \mathcal{G}_h(z_i, z) }
+ \Ord(q^2)\, ,
\label{S1}
\\
S_2 = 
\int \d^2 z  \, 
&\left [ 1 + \sum_{i=1}^N  \frac{\a'}{2} k_i q  \mathcal{G}_h(z_i, z) 
\right ]
 \frac{\e \cdot \bar{\e}}{2}   
\  \omega_I(z) (2 \pi \, \Imt )_{IJ}^{-1} \, \bar{\omega}_J(\bar{z})
+ \Ord(q^2)\, .
\label{S2}
}
Notice that the integrand of $S_1$ is expanded through order $q^2$, since the integration can bring down one order of $q$. We will only keep terms through order $q$ after integration.

Let us first notice that the first integral in $S_2$ immediately follows from the \emph{Riemann Bilinear Identity}; 
since the abelian forms are closed holomorphic
forms we have
\ea{
\int d^2 z\  \omega_I (z) \bar{\omega}_J (\bar{z}) 
=- i \sum_{\sigma = 1}^h \left [ \oint_{a_\sigma} \omega_I \oint_{b_\sigma} \bar{\omega}_J - \oint_{b_\sigma} \omega_I \oint_{a_\sigma} \bar{\omega}_J \right ]
= 4\pi (2 \pi {\rm Im} \tau_{I J} )\, ,
\label{RBI}
} 
where the definition of the abelian cycles, outlined in \App{Schottky}, was used.
It follows that
\ea{
S_2 = (\e \cdot \bar{\e}) \left [ 2\pi h  + S_2^{(1)} + \Ord (q^2) \right ]\, ,
\label{eq:S2}
}
where $S_2^{(1)}$ denotes the contribution at order $q$, given by the integral
\ea{
S_2^{(1)} = 
\sum_{i=1}^N  \frac{\a'}{4} k_i q
\int \d^2 z  \, 
  \mathcal{G}_h(z_i, z) 
 \ \omega_I(z) (2 \pi \, \Imt )_{IJ}^{-1} \, \bar{\omega}_J(\bar{z})\, .
 \label{S21}
}
This quantity is discussed in App. \ref{X}. 
At one loop it turns out that the integration gives an expression independent of $z_i$, 
and hence the total expression vanishes as a consequence of momentum conservation,
i.e. $\sum_{i=1}^N k_i \cdot q = - q^2 = 0$. At the multiloop level, we have not been able to evaluate this integral.
We notice for future studies that if the one loop result holds at multiloops, it is not necessary to calculate the integral explicitly, but only to show that the integral is independent of $z_i$.

To compute $S_1$ let us first remark that we can restrict to the case where the soft state is symmetrically polarized, 
since the amplitude of one Kalb-Ramond state and $N$ closed tachyons is anyways zero because of {world-sheet} parity conservation%
\footnote{%
More precisely,
world-sheet parity $\Omega$, which is a symmetry of the closed bosonic string, leaves invariant the vertex operators of the tachyon, dilaton and graviton, while changing sign of the Kalb-Ramond vertex operator.}.
In the case where the soft state is symmetrically polarized, 
it remarkably turns out that $S_1$ is computable through order $q$ by using only the identities in \Eq{d2G} and \Eq{Intd2G}, i.e. independent of the explicit form of $\G_h$. 
We leave the details of this important result to the \App{CalS1} and here quote the final expression:
\ea{
S_1 = &\ 2
\pi \varepsilon_{q\mu\nu}^S \sum_{i=1}^N\frac{k_i^\mu k_i^\nu}{k_iq}
+\ 2 \pi  \varepsilon_{q\mu\nu}^S \frac{\a'}{2}  \sum_{i\neq j} \left [
 \frac{k_i^\mu k_i^\nu}{k_iq} (k_jq)  \G_h(z_j,\,z_i) - 
 k_i^\mu k_j^\nu \G_h(z_j,\,z_i)
 \right ]
\nonumber \\[2mm]
&
+ 2 \pi  \varepsilon_{q\mu\nu}^S \frac{1}{2}  \left(\frac{\alpha'}{2} \right)^2 \sum_{i\neq j,l} \Bigg [
k_i^\mu k_i^\nu \frac{ (k_jq)(k_lq)    }{qk_i}
+ k_j^\mu k_l^\nu (k_i q) 
\nonumber \\[2mm]
& \hspace{4cm}
-  k_i^\mu k_l^\nu (k_jq)- k_i^\mu k_j^\nu (k_l q) 
\Bigg ]\G_h(z_i,\,z_l) \G_h(z_i,\,z_j)  + \Ord(q^2)\, .
\label{S1explicit}
}
Noticeably, this result is formally equal to the tree-level result found in {Eq.~(2.5) of Ref.~\cite{DiVecchia:2015oba}, upon inserting the corresponding tree-level Green function. 
It follows that it is reproduced by the same soft theorem as valid at tree-level, which we can now easily check: The first term is just the Weinberg soft theorem, when multiplied with $N_0$, which immediately factorizes out of $M_n$, since it is independent on the Koba-Nielsen variables. The subleading terms should be reproduced by the following operation:
\ea{
-i  \kappa_D\, \varepsilon_{q\mu\nu}^S \sum_{i=1}^N\frac{k_i^\mu q_\rho L_i^{\nu \rho}}{k_i q} M_N^{(h)} &=
\kappa_D\, \varepsilon_{q\mu\nu}^S \sum_{i=1}^N\frac{k_i^\mu q_\rho }{k_i q} \left (k_i^\nu \partial_{k_i}^\rho - k_i^\rho \partial_{k_i}^\nu  \right ) M_N^{(h)}
 \nonumber \\
 &=
\kappa_D\, M_N^{(h)} \ast \varepsilon_{q\mu\nu}^S  \sum_{i\neq j }^N\frac{k_i^\mu q_\rho }{k_i q}   \left [
k_i^\nu k_j^\rho - k_i^\rho k_j^\nu \right ] \G_h(z_i, z_j)
 \nonumber \\
 &=
\kappa_D\, M_N^{(h)} \ast \varepsilon_{q\mu\nu}^S   \frac{\a'}{2} \sum_{i\neq j }^N\left [ \frac{k_i^\mu k_i^\nu }{k_i q} (k_j q) - 
 k_i^\mu k_j^\nu \right ]\G_h(z_i, z_j)\, ,
 }
 which is exactly equal to the subleading soft term in \Eq{S1explicit} when multiplied with $N_0$.
 
 Finally, the subsubleading tree-level soft operator reads~\cite{DiVecchia:2015oba}:
 \ea{
 &- \kappa_D\, \varepsilon_{q\mu\nu}^S \sum_{i=1}^N\frac{q_\rho q_\sigma }{2k_i q}: L_i^{\mu \rho} L_i^{\nu \sigma} : M_N^{(h)}
 \nonumber \\
 &
 = 
\kappa_D\, \varepsilon_{q\mu\nu}^S \sum_{i=1}^N\frac{q_\rho q_\sigma }{2k_i q} 
 \left ( k_i^\mu k_i^\nu \partial_{k_i}^\rho\partial_{k_i}^\sigma
 + k_i^\rho k_i^\sigma \partial_{k_i}^\mu\partial_{k_i}^\nu
 -k_i^\mu k_i^\sigma \partial_{k_i}^\nu\partial_{k_i}^\rho
 -k_i^\nu k_i^\rho \partial_{k_i}^\mu\partial_{k_i}^\sigma
 \right ) M_N^{(h)}
  \nonumber \\
 &
 =  M_N^{(h)} \ast \frac{\kappa_D}{2} \left (\frac{\a'}{2} \right )^2 \varepsilon_{q\mu\nu}^S \sum_{i=1} \sum_{j,l\neq i} \Bigg\{
 k_i^\mu k_i^\nu \frac{ (k_jq)(k_lq)    }{qk_i}
+ k_j^\mu k_l^\nu (k_i q) 
\nonumber \\
& \hspace{5cm}
-  k_i^\mu k_l^\nu (k_jq)- k_i^\mu k_j^\nu (k_l q) 
\Bigg \}\G_h(z_i,\,z_l) \G_h(z_i,\,z_j)\, ,
 }
 which is exactly equal to the subsubleading term in \Eq{S1explicit} when multiplied with $N_0$.
The term containing two L  in the subsubleading soft operator above is  \emph{normal ordered}, hence the $: {} :$ notation, meaning that the operator $L$ on the left acts on everything on its right except on the other $L$.
The normal ordering is irrelevant for the graviton, but is important for getting the right behavior of the dilaton.

To summarize, we have found that the $h$-loop soft behavior of the graviton and dilaton when scattering with $N$ tachyons in the bosonic string can be written as:
\ea{
M_{N;1}^{(h)} (k_i; q) = 
&\kappa_D \, \varepsilon_{q\mu\nu}^S \sum_{i=1}^N\Bigg [ \frac{k_i^\mu k_i^\nu}{k_iq}
-i \frac{k_i^\mu q_\rho}{k_i q}  L_i^{\nu \rho}
-\frac{q_\rho q_\sigma }{2k_i q}: L_i^{\mu \rho} L_i^{\nu \sigma} : 
+  h \,  \eta^{\mu \nu}
\Bigg] M_N^{(h)}\\
&+
(\epsilon \cdot  \bar{\epsilon} ) \, \frac{\kappa_D}{2 \pi} \left ( M_N^{(h)} \ast S_2^{(1)}\right )
 + \Ord(q^2)
\label{SoftTheoremOrderh}
}
The last term is zero at least at tree-level and at one loop.

\subsection{All-loop graviton soft theorem}

The full amplitude of one graviton and $N$ closed tachyons is 
\ea{
\mathcal{M}_{N;g} (k_i; q) = \sum_{h=0}^\infty M_{N;1}^{(h)} (k_i; q) \Big |_{\varepsilon_{q\mu \nu}^S = \varepsilon_{\mu \nu}^g}\, .
}

Since the polarization tensor of the graviton is traceless, $\varepsilon_{\mu \nu}^g \eta^{\mu \nu} = 0$,
it simply follows that the full all-loop graviton soft behavior is given by the soft theorem:
\ea{
\mathcal{M}_{N;g} (k_i; q) = 
\kappa_D \, \varepsilon_{q\mu\nu}^S \sum_{i=1}^N\Bigg [ \frac{k_i^\mu k_i^\nu}{k_iq}
-i \frac{k_i^\mu q_\rho}{k_i q}  L_i^{\nu \rho}
-\frac{q_\rho q_\sigma }{2k_i q}: L_i^{\mu \rho} L_i^{\nu \sigma} :
\Bigg] \mathcal{M}_N(k_i)  + \Ord(q^2)\, ,
}
where $\mathcal{M}_N{(h)}$ is the full all-loop amplitude of $N$ closed string tachyons.
These amplitudes are, of course, plagued by infrared divergences, and we are here tacitly assuming
that the soft limit is taken before any infrared divergent limit. This issue will be discussed in a subsequent section.

\subsection{All-loop dilaton soft theorem}
The full amplitude of one dilaton and $N$ closed tachyons is 
\ea{
\mathcal{M}_{N;\phi} (k_i; q) = \sum_{h=0}^\infty M_{N;1}^{(h)} (k_i; q) \Big |_{\varepsilon_{q\mu \nu}^S = \varepsilon_{\mu \nu}^\phi}\, ,
}
where $\varepsilon_{\mu \nu}^\phi = \frac{1}{\sqrt{D-2}} (\eta_{\mu \nu} - q_\mu \bar{q}_\nu - q_\nu \bar{q}_\mu)$
with $\bar{q}^2 = 0$ and $q \cdot \bar{q} = 1$. 
After contracting with this projection tensor, the $h$-loop soft behavior for the dilaton becomes:
\ea{
M_{N;\phi}^{(h)} (k_i; q) = &
\frac{\kappa_D}{ \sqrt{D-2} }  \left[ - \sum_{i=1}^N \frac{m^2}{k_i q} {\rm e}^{q \partial_{k_i}}  + 2 - \sum_{i=1}^N\hat{D}_i
+ h  (D-2)
+  q_\mu \sum_{i=1}^N {\hat{K}}_{i}^{ \mu}
   \right]  M_N^{(h)}    
   \nonumber \\
   &
 +   \frac{\kappa_D}{2 \pi} \sqrt{D-2} \, \left ( M_N^{(h)} \ast S_2^{(1)}\right )
   + \Ord(q^2) \, , 
\label{dilasoftfin}
}
where in the case of tachyons $m^2 =- \frac{4}{\alpha'}$ and 
\ea{
\hat{D}_i = k_{i} \cdot  
\frac{\partial}{\partial k_{i}}  \, , 
\qquad 
 \hat{K}_{i}^{\mu} = \frac{1}{2} k_{i}^{ \mu} \frac{\partial^2}{\partial k_{i\nu} \partial k_i^\nu}  -k_{i}^{\rho} \frac{\partial^2}{ \partial k_{i}^{\rho} \partial k_{i\mu}}  \, ,
\label{hatDhatKmu1}
}
which are the momentum space generators of space-time dilatations and special conformal transformations.
The last term in \Eq{dilasoftfin}, which we have not been able to evaluate at the multiloop level, is of order $q$ and could potentially break the factorizing soft behavior of the amplitude at this order.
At one-loop order, however, it turns out as explained earlier (see also \App{X}) that
it vanishes. Thus at least at one loop we have a soft theorem for the dilaton through subsubleading order, reading:
\ea{
M_{N;\phi}^{(1)} (k_i; q) = 
\frac{\kappa_D}{ \sqrt{D-2} }  \left[ - \sum_{i=1}^N \frac{m^2}{k_i q} {\rm e}^{q \partial_{k_i}}  + D - \sum_{i=1}^N\hat{D}_i
+  q_\mu \sum_{i=1}^N {\hat{K}}_{i}^{ \mu}
   \right]  M_N^{(1)}    + \Ord(q^2) \, .
\label{dilasoftfin1loop}
}

We cannot immediately write the all-loop soft behavior in this case, because of the explicit dependence on $h$ in the dilaton soft operator. 
However, notice that the $h$-loop amplitude has the following scaling property
\ea{
M_N^{(h)}  = \sqrt{\alpha'}^{(2-D)h -2} \kappa_D^{2(h-1)+N} {F} \left(\sqrt{\alpha'} k_i , R/\sqrt{\alpha'} \right ) \, ,
}
where ${F}$ is a dimensionless function parametrizing the amplitude, {and we recall that $R$ is denoting the compactification radii, which enter only for $h>0$.}
The gravitational constant is given in terms of $\alpha'$, $g_s$, the string coupling constant, {and $R$}, as follows {(see App.~\ref{App:kappaD} for a derivation):
\ea{
\kappa_D = (2\pi)^{\frac{D-3}{2}} \,\sqrt{2^{-9}}\, g_s\, \sqrt{\alpha'}^{\frac{D-2}{2}}
\left (\frac{\sqrt{\alpha'}}{ R}\right )^{\frac{26-D}{2}}
\, .
}
}
From these expressions we deduce that 
\ea{
\left [2 - \sum_{i=1}^N k_i \cdot \frac{\partial}{\partial {k_i}} +(D-2)h  \right ] M_N^{(h)} = 
\left [\frac{D-2}{2} g_s \frac{\partial}{\partial g_s} - \sqrt{\alpha'}\frac{\partial}{\partial \sqrt{\alpha'} } 
{- R \frac{\partial}{\partial R}} \right ] M_N^{(h)} 
\, .
}
The left-hand side is nothing but the subleading $h$-loop soft dilaton operator. It is exactly reproduced by the operator on the right-hand side, which is $h$-independent. Notice also that the operator on the right-hand side leaves $\kappa_D$ invariant.
{For $h=0$, the $R$-dependence is only in $\kappa_D$, why in that case one can rewrite
the soft operator in terms of a $D$-dimensional string coupling constant and $\alpha'$; see \Eq{tree-level-soft-operator}, which explains previous tree-level result where the additional $R$-operator did not appear.}
Slightly more explicit considerations on this soft operator are offered in App.~\ref{App:kappaD}, where a general $26-D$-dimensional toroidal compactification is considered.

Now we can sum all loop contributions to form the full amplitude on the right-hand side of the soft theorem, yielding 
through subleading order
\ea{
\mathcal{M}_{N;\phi} (k_i; q) = 
\frac{\kappa_D}{ \sqrt{D-2} } 
\left[ - \sum_{i=1}^N \frac{m^2}{k_i q} {\rm e}^{q \partial_{k_i}}  
+ \frac{D-2}{2} g_s \frac{\partial}{\partial g_s} - \sqrt{\alpha'} \frac{\partial}{\partial \sqrt{\alpha'}}   
{- R \frac{\partial}{\partial R}} \right] \mathcal{M}_N(k_i) + \Ord(q) \, .
}
This gives the all-loop dilaton soft behavior through subleading order when scattering with $N$ closed tachyons.
The same discussion about infrared divergences mentioned in the graviton case applies also here.

\section{Infrared Divergences}
\setcounter{equation}{0}
\label{IRD}

In the previous section we have shown that the graviton and dilaton satisfy soft factorization theorems at $h$ loops,
but we have not taken into account that actually the multiloop amplitudes  in the bosonic string are infrared divergent. In this section we discuss how to treat them
preserving the results that we have already obtained.  

In the bosonic string, where we compactify 
$(26 -D)$ dimensions leaving  $D$ non-compact dimensions, we have two kinds of infrared divergences: The first kind, appearing for any value of $D$, arises due to the fact that the bosonic string has a tachyon and a non-vanishing dilaton tadpole.  {As we shall see, this kind of} infrared divergence  does not depend on the number of external legs. {When massless states are involved, another kind of infrared divergence may appear when we approach low values of non-compact dimensions, for instance when $D=4$, which from field theory are known as soft and collinear divergences.
These instead depend on the number of external legs, which we shall also briefly discuss from the string theory perspective.}

{String amplitudes involving external tachyons, or more generally involving massive external states are also plagued by additional divergences that require mass-renormalization. As explained in Ref.~\cite{WeinbergT85}, these divergences can be regularized by not allowing the Koba-Nielsen variables to get too close to each other in certain configurations  (see also Ref.~\cite{Renormalization}, and the recent progress in Ref.~\cite{Sen-Renormalization}). 
But, since they depend only on the number of external massive legs, we do not expect that they will modify the soft operator.}

Let us discuss the first kind of infrared divergences in the simplified case of the $N$-tachyon amplitude at one loop, {which is given by Eq.~(\ref{3.1b}) for the case $h=1$.
After explicitly deriving the $h=1$ expressions, setting $z_i=e^{2\pi i\nu_i}$,
and defining $\nu_{ij} = \nu_i - \nu_j$,
the one-loop $N$-tachyon amplitude reads:
\begin{eqnarray}
T_N^{(1)} = &&C_1 N_0^N \int_{{\cal{F}}} d^2 \tau \, \, \mu (\tau, {\bar{\tau}} ) \prod_{i=1}^{N-1} 
\left[\int d^2 \nu_i  \right] \nonumber \\
 && \times  \prod_{i < j} \Big| \frac{\sin \pi \nu_{ij}}{\pi} \prod_{n=1}^\infty
\frac{(1- \mk^n {\rm e}^{2\pi i \nu_{ij}} ) ( 1- \mk^n {\rm e}^{-2\pi i \nu_{ij}} )}{(1-\mk^n )^2}
{\rm e}^{-\pi \frac{({\rm Im}\,\nu_{ij})^2}{\Imt}} \Big|^{ {\alpha'} k_i k_j} \, ,
\label{IR1}
\end{eqnarray}
where we have set $\eta=0$, $\xi=\infty$, $\nu_N =0$,  $\mk = {\rm e}^{2\pi i \tau}$.
$\mathcal{F}$ denotes the fundamental integration region of $\tau$, and
\begin{eqnarray}
\mu (\tau , {\bar{\tau}}) =(2\pi)^2 {\rm e}^{4 \pi \Imt} 
\prod_{n=1}^{\infty}\left[ \frac{1}{|1 - {\rm e}^{2\pi i \tau n} |^{48}}\right] 
\frac{ (F(\tau , {\bar{\tau}}))^{26-D}}{ (\Imt)^{D/2}}  \, .
\label{IR40}
\end{eqnarray} 

We now consider the region of the moduli space where all $\nu_i$ are very close to each other and to $\nu_N=0$. This can be done by introducing the variables $\eta_i , \varepsilon, \phi$ as follows:
\begin{eqnarray}
{\rm e}^{i \phi} \varepsilon \eta_i = \nu_i \,\,\,, \,\, i=1 \dots N-2~~;~~ 
 \varepsilon {\rm e}^{i \phi} = \nu_{N-1}~~;~~ \eta_{N-1} =1 \, .
\label{IR29}
\end{eqnarray}
By using these variables and  keeping only the terms divergent for  $\varepsilon \rightarrow 0$
we find
\begin{eqnarray}
T_n^{(1)}=&& C_1 N_0^N \int_{\cal{F}} d^2  \tau \mu (\tau, {\bar{\tau}} ) 
 \prod_{i=1}^{N-2} \int d^2 \eta_i    \int_0^{2\pi} d \phi \int_0^1  \frac{d \varepsilon}{\varepsilon^{3 - \frac{\alpha'}{2} p^2}} \prod_{i<j}  | \eta_{ij} 
|^{\alpha' k_i k_j} \nonumber \\
 &&  \times  \left[ 1  - \alpha' \sum_{i<j}  k_i k_j \frac{\pi \varepsilon^2}{\Imt} \left( \sin \phi \, \Re ( \eta_{ij}) + \cos \phi \, {\rm Im} ( \eta_{ij}) \right)^2  + \Ord(\varepsilon^4) \right]  \, ,
\label{IR50}
\end{eqnarray}
where by $\phi$-integration a second $\varepsilon^{-1}$-term was removed, and where the divergent terms
for $\varepsilon \sim 0$ were regularized by the substitution $- 3 \rightarrow -3 + \frac{\alpha'}{2} p^2$}. The upper limit in the integral was set to $\varepsilon=1$, 
since we are interested in the behavior near $\varepsilon \sim 0$.
The integrals over $\phi$ and $\varepsilon$ can be performed and one gets:
\begin{eqnarray}
T_N^{(1)}= && C_1 N_0^N \int_{\cal{F}} d^2  \tau \mu (\tau, {\bar{\tau}} )  \prod_{i=1}^{N-2} \int d^2 \eta_i   \prod_{i<j}  | \eta_{ij}    |^{\alpha' k_i k_j}
 \nonumber \\
&& \times  
\left[ - \frac{2\pi}{2 - \frac{\alpha'}{2} p^2} + \frac{2\pi}{p^2} \frac{\pi }{\Imt} \sum_{i<j} k_i k_j  \eta_i   {\bar{\eta}}_j    \right] + \cdots\, ,
\label{IR91}
\end{eqnarray}
where the dots denote terms that are regular for $p^2 \rightarrow 0$.
The first term in the last line corresponds to the regularized tachyon contribution, while the second term, corresponding to the dilaton contribution,  is divergent when $p^2 \rightarrow 0$. 
The second term in the square bracket has been obtained by using the identity:
\begin{eqnarray}
\sum_{i<j} k_i k_j |\eta_i - \eta_j |^2  = \frac{1}{2} \sum_{i,j=1}^N  k_i k_j |\eta_i - \eta_j |^2 
 = -  \sum_{i, j =1}^N k_i k_j  \eta_i {\bar{\eta}}_j\, ,
\label{IR35a}
\end{eqnarray}
which follows from momentum conservation for $p \sim 0$.

Finally, we  recognise that the coefficients of the two poles are the tree-level amplitude with $(N+1)$ tachyons and that with $N$ tachyons and one dilaton that we rewrite here:
\begin{eqnarray}
&&T_{(N+1){\rm  tach}} = C_0 N_0^{N+1}  \prod_{i=1}^{N-2} \int d^2 \eta_i \prod_{i<j =1}^{N} 
 | \eta_i - \eta_j |^{\alpha' k_i k_j}\, ,
\nonumber \\
&& T_{{N}{\rm tach}+ 1{\rm dil}} =  C_0 N_0^{N+1} \frac{\alpha'}{2} \prod_{i=1}^{N-2} \int d^2 \eta_i \prod_{i<j =1}^{N} | \eta_i - \eta_j |^{\alpha' k_i k_j} \sum_{i,j=1}^N k_i^\mu k_j^\nu \eta_i {\bar{\eta}}_j\, .
\label{IR76}
\end{eqnarray}
Using them  and the fact that $p \sim 0$ we can rewrite Eq.~(\ref{IR35a}) as follows~\cite{Ademollo:1975pf,Shapiro:1975cz}:
\begin{eqnarray}
T_N^{(1)} = \frac{C_1}{C_0 N_0} \int_{{\cal{F}}} d^ 2 \tau \mu (\tau , {\bar{\tau}}) \left[
- \frac{2\pi T_{(N+1)\rm tach}}{2- \frac{\alpha'}{2} p^2}  + \frac{(2\pi)^2}{\alpha' p^2 \Imt} T_{N {\rm tach}+1{\rm dil}}\right]+ \cdots\, .
\label{IR39}
\end{eqnarray}
In conclusion, we get  the first term in the square bracket in Eq.~(\ref{IR39}) that has the propagator of the  tachyon  times the amplitude with $(N+1)$ tachyons with one of them connected to the closed string tadpole that has to be regularized for large values of $\Imt$. The introduction of the finite momentum $p$ has regularized the contribution of the tachyon propagator. The second term in the square bracket has the pole of the massless dilaton  attached to the closed string tadpole, times the amplitude with $N$ tachyons and one dilaton. This term is divergent when $p \rightarrow 0$, but can be regularized keeping $p \neq 0$.  

In conclusion, we have shown that, in a certain region of the moduli space the bosonic string exhibits the divergence due to the dilaton tadpole. We have regularized it by introducing an infrared cutoff $p$. As already mentioned, another cutoff for $ \Imt \rightarrow \infty$ should also be introduced to regularize the closed string  tadpole. These divergences are, however, present  in both  the amplitude with $(N+1)$ particles and in that with $N$ particles that appear in the soft theorem.  Therefore, by regularizing both of them in the same way with an infrared  cutoff,   
the soft theorems found in the previous section are still satisfied.

If we take the number of non-compact directions down to $D=4$, then one gets additional infrared divergences. They are obtained in the limit $ \Imt \rightarrow \infty$ and are discussed in detail in Ref.~\cite{Green:NPB198} for the closed superstring for  amplitudes with massless states.  They 
correspond to 
{the soft infrared divergences}
that one also finds in field theory that, depending  on the number of external legs, prevents the soft theorem, found at the tree level, to be also valid at loop level~\cite{Bern:2014oka}.
{In field theory, one also encounter collinear divergences in certain Feynman loop diagrams of massless amplitudes.
However, it is a time-honored result due to Weinberg~\cite{Weinberg:1965nx}, which has since been proven to all orders in perturbation theory~\cite{1109.0270}, that such divergence do not appear in the full amplitude involving gravitons and other massless states.}
We will not consider these divergences in more detail in this paper and refer to Ref.~\cite{Green:NPB198} for the interested reader (see also Ref.~\cite{Laddha:2018myi} for a discussion on these divergences in the classical limit).

In conclusion, the multiloop soft theorem found in this paper is
valid when the number of non-compact dimensions is greater than four.

\clearpage

\section{Conclusions}
\label{conclusion}
\setcounter{equation}{0}
 
{In this paper we have   extended to multiloops in the bosonic string the soft theorems 
for the graviton  and the dilaton.}
{This has been done by computing the amplitude with one  graviton/dilaton and $N$ tachyons and by explicitly studying its behaviour in the limit where the massless states carry  low momentum. The main ingredient used  to compute this  amplitude to  an  arbitrary order of the perturbative expansion has been the $h$-loop $N$-Reggeon vertex}. This is an operator, constructed in the 80s, that when applied on an arbitrary number of  external states, not necessarily on-shell, provides  the scattering amplitudes corresponding to  the  external states taken in consideration.

The amplitude that we have computed, as any { loop  amplitude in the bosonic string},  suffers from infrared (IR) divergences. They appear in different corners  of the moduli space and in principle  may modify the tree-level soft operators. 
One corner of moduli space that leads to these IR divergences 
corresponds to a world-sheet topology where  one or more loops, tadpoles, are connected by long tubes to a sphere where all the external states are inserted.
The integration over the modulus parametrizing  the length of this tube gives rise to divergences which are due to the exchange  between the loops  and the sphere of   tachyons and on-shell dilatons. We have regularized these divergences 
by introducing a finite momentum for the intermediate states. Furthermore we have regularized the closed string tadpole by introducing a cut-off for large values of the string moduli.   These regularizations, being independent on the number of external states,  do not  affect soft theorems.
  
Another  class of IR divergences appears  when   string theories are  compactified down to four dimensions. They  have been 
studied in literature\cite{Green:NPB198} and the main feature is that {they} depend  on the number of the external states and therefore affect the soft theorems. {They can be avoided by compactifying string amplitudes down to a space-time dimension $D>4$~\cite{Laddha:2018myi}.}
Under such a limitation and for amplitudes with one soft graviton/dilaton and $N$ hard tachyons the main result of our analysis  is  that the soft graviton  behaviour at loop level coincides with the one at the tree level. {This happens because the loop amplitudes have the same form, in terms of the Green function, as the tree-level ones, except for an extra contribution in the case of the dilaton which is importantly there to ensure the right scaling relation at every loop order. 

{It is by now well established} that tree-level soft theorems of gauge  and gravitational fields are a consequence of the gauge symmetries underlying the theory. The IR behaviour of an amplitude, $M_{N+1} = \varepsilon_{\mu \nu} M^{\mu\nu}_{N+1}$, with a  graviton carrying low momentum $q$ and $N$ hard particles is fully  determined, up to the order $q$, by  the gauge invariance conditions:
\begin{eqnarray}
q_\mu \left(M^{\mu\nu}_{N+1}(q,\,\{k_i\})-f(q,\,\{k_i\})\eta^{\mu\nu}\right)=q_\nu \left(M^{\mu\nu}_{N+1}(q,\,\{k_i\})-f(q,\,\{k_i\})\eta^{\mu\nu}\right)=0\, .\label{extra}
\end{eqnarray}
Here $f(q,\,\{k_i\})$ is an arbitrary function of the momenta that contributes  only in the case of a soft dilaton. Its  arbitrariness prevent us to get the full soft behaviour of the dilaton from gauge invariance. 
In string theory this quantity can be explicitly computed and from Ref.~\cite{Ademollo:1975pf} it is well known that it differs from  zero in the presence of massless open string states. In this paper we have  seen  that it also differs from zero at the loop level of only closed strings but that the soft behaviour of  multiloop amplitudes with only closed string states can be obtained from  gauge invariance  by taking this function equal to the number of handles, $h$, of the Riemann surface, at least through subleading order. This choice does not affect the tree-level gauge conditions but the loop ones are modified in such a way to reproduce exactly the extra contribution given in Eq.~(\ref{eq:S2}). 
At the level of the full amplitude, the extra contribution at loop-level singles out the scaling operator
\ea{
\frac{D-2}{2} g_s \frac{\partial}{\partial g_s} - \sqrt{\alpha'}\frac{\partial}{\partial \sqrt{\alpha'} } 
{- R \frac{\partial }{\partial R}}
\, ,
}
as the proper multiloop dilatation operator determining the order $q^0$ soft behavior of the dilaton.
{Here $R$ denotes collectively the compactification radii.}

It is not clear yet if the  tree-level universality of the soft dilaton theorem, 
i.e. its independence of the string theory considered, extends to loops. Indeed this paper, in a special case, shows that the multiloop soft dilaton operator, when written in terms of the string slope, string coupling constant, {and compactification parameters}, is the same as at tree-level at least through subleading order, and at one loop we could show that they are the same even through subsubleading order. It remains an open problem to understand whether the subsubleading one loop behavior extends to all loops, and it would be very interesting to explore its universality at loop level 
in other string theories and for  arbitrary hard states.

\subsection*{Acknowledgments} \vspace{-3mm}

We thank Igor Pesando and Rodolfo Russo  for many useful discussions. We additionally thank Rodolfo Russo, and Stefano Sciuto for a critical reading of the manuscript.
{We also would like to thank the referee of this paper and Massimo Bianchi for, respectively in the referee report and in a seminar, pointing out to us that the dependence on the compactification radii was not fully taken into account in the formulation of the dilaton soft theorem that appeared in the first version of the paper.}

\appendix

\clearpage
\section{On the gravitational coupling constant in string theory}
\label{App:kappaD}
\setcounter{equation}{0}

For both the  bosonic string, in the critical $d=26$ dimensions,  and the superstring, in   $d=10$, one can derive the following formula for the  $Dp$-brane  tension $T_p$\cite{PolchinskiI,PolchinskiII}:
\begin{eqnarray}
T_p = \frac{\sqrt{\pi}}{2^{\frac{d-10}{4}}} (2\pi \sqrt{\alpha'})^{\frac{d}{2} -p-2 }~~;~~\tau_p \equiv \frac{T_p}{\kappa_d}
\label{P1}
\end{eqnarray}
where $\tau_p$ is the physical tension of a Dp-brane. In Eq. (13.3.20) of Ref.\cite{PolchinskiII} it is proposed that the string coupling constant $g_s$ is given by the ratio of the fundamental string tension $\tau_{F1} = \frac{1}{2\pi \alpha'}$ and the tension $\tau_{D1}$ of the D1-brane:
\begin{eqnarray}
\frac{\tau_{F1}}{\tau_{D1}}= g_s = \frac{\kappa_d}{2\pi \alpha'}\frac{ 2^\frac{d-10}{4}}{\sqrt{\pi} (2\pi \sqrt{\alpha'})^{\frac{d}{2} -3}}
\label{P2}
\end{eqnarray}
It implies
\begin{eqnarray}
2 \kappa_d^2 = \frac{g_s^2  (2\pi)^{d-3} (\sqrt{\alpha'})^{d-2} }{2^\frac{d-10}{2}} 
\label{P3}
\end{eqnarray}
that for the bosonic and superstring respectively gives:
\begin{eqnarray}
2 \kappa^2_{26} = 2^{-8} g_s^2 (2\pi)^{23} (\alpha')^{12}~~;~~
2 \kappa_{10}^2 = g_s^2 (2\pi)^7 (\alpha')^4~~
\label{P4}
\end{eqnarray}
We consider in the following the compactification of the effective action of the  bosonic string theory, where $d=26$. The analysis, however, can be equally performed in the $d=10$ superstring theory case.
The  terms of  the effective  action of the bosonic string  theory,  in the string frame, relevant for the forthcoming discussion are:
\begin{eqnarray}
S=\frac{{\rm e}^{-2 {\hat{\phi}}_0}}{2 \hat{\kappa}^2} \int d^{26} \hat{x} \sqrt{|\hat{G}|} \,{\rm e}^{-2 {\hat{\phi}}} \left( \hat{R}(\hat{x}) + \dots \right)~~;~~{\hat{\phi}} = \phi_{26} - {\hat{\phi}}_0
\label{P8}
\end{eqnarray}
where
\begin{eqnarray}
2 \hat{\kappa}^2  = 2^{-8} (2\pi)^{23} (\alpha' )^{12} = 2 (\kappa_{26}/g_s)^2
\label{P8a}
\end{eqnarray}
Compactifying  $26-D$-dimensions, the $26$-dimensional  Lorentz group $SO(1,25)$ is broken to $SO(1,D-1)\times SO(26-D)$.  We hence decompose the space-time coordinates into $\hat{x}^M=(x^\mu, y^m)$ with $\mu=0\dots D-1$ and $m=D \dots 25$. 
Under the assumption that the $26$-dimensional metric, $\hat{G}_{MN}$, admits the following block-diagonal form
\begin{eqnarray}
\hat{G}_{MN}=\left(\begin{array}{cc}
            {g}_{\mu\nu}(x^\mu)&0\\
            0& G_{mn}(x^\mu)\end{array}\right)~~;~~M,N=0\dots 25
 \label{L3}
\end{eqnarray} 
the low-energy string action can be reduced to an effective $D$-dimensional action, $S_D$, as follows: First, under \Eq{L3}, the action reduces to
\begin{eqnarray}
S_D=\frac{{\rm e}^{-2 {\hat{\phi}}_0}}{2 \hat{\kappa}^2} \int d^{D} x \sqrt{|g|} \int_0^{2\pi \sqrt{\alpha'}} d^{26-D} y \sqrt{G} \,e^{-2\hat{\phi}}\left( \hat{R}(x) + \dots \right)
\label{P9}
\end{eqnarray}
Next, introduce the $D$-dimensional dilaton, $\phi(x)$, through:
\begin{eqnarray}
{\rm e}^{-2 {\hat{\phi}_0}} \sqrt{G} \,e^{-2\hat{\phi}}
\equiv {\rm e}^{-2\phi} {\rm e}^{-2 \phi_0}
\label{P10}
\end{eqnarray}
where
\begin{eqnarray}
 {\rm e}^{-2\phi} = {\rm e}^{-2 {\hat{\phi}}} \frac{\sqrt{G}}{\sqrt{G_0}} ~~;~~{\rm e}^{-2 \phi_0} ={\rm e}^{-2 {\hat{\phi}}_0} \sqrt{G_0}  \, , 
\label{P11}
\end{eqnarray}
and $G_0$ denotes the vacuum expectation value of the compact metric.
Taking the compact manifold to be the product of $26-D$ circles with radii $R_m$, we get
\begin{eqnarray}
\sqrt{G_0} =  \frac{\prod_{m=1}^{26-D}R_m}{(\sqrt{\alpha'})^{26-D}}
\label{P12}
\end{eqnarray}
With these ingredients, Eq. (\ref{P9}) becomes:
\begin{eqnarray}
S_D && = \frac{(2\pi {\sqrt{\alpha'})^{26-D}}{\rm e}^{-2 {{\phi}}_0}}{2 \hat{\kappa}^2} \int d^{D} x \sqrt{|g|}  {\rm e}^{- 2 \phi}\left( \hat{R}(x) + \dots \right) \nonumber \\
&& = \frac{1}{2 \kappa_D^2} \int d^{D} x \sqrt{|g|}  {\rm e}^{- 2 \phi}\left( \hat{R}(x) + \dots \right)
\label{P13}
\end{eqnarray}
where we note that $\hat{R}(x)$ is the Ricci scalar computed from the $26$-dimensional metric, \Eq{L3}, which, however, only depends on the $D$-dimensional noncompact coordinates $x^\mu$.
It follows that
\begin{eqnarray}
\frac{1}{2 \kappa_D^2} 
=
 \frac{(2\pi {\sqrt{\alpha'})^{26-D}}{\rm e}^{-2 {\hat{\phi}}_0} \sqrt{G_0}}{2 \hat{\kappa}^2}  
 = \left (\frac{g_s^2}{{\rm e}^{2{\hat{\phi}}_0}}\right )
 \frac{ \prod_{m=1}^{26-D}(2\pi R_m)}{2 \kappa_{26}^2}   \, .
\label{P14}
\end{eqnarray}
Following conventions, we set $g_s = {\rm e}^{\hat{\phi}_0}$, whereby the prefactor above is simply one.
 Inserting the explicit expression from Eq.~(\ref{P4}) for $\kappa_{26}$ we get
{
\begin{eqnarray}
\kappa_D 
= \frac{g_s (2\pi)^{23/2} (\alpha')^6}{\sqrt{2}^9 \prod_{m=1}^{26-D}\sqrt{2\pi R_m}}
=(2\pi)^{\frac{D-3}{2}} \,\sqrt{2^{-9}}\, g_s\, \sqrt{\alpha'}^{\frac{D-2}{2}}
\prod_{m=1}^{26-D}\left (\frac{\sqrt{\alpha'}}{ R_m}\right )^{\frac{1}{2}}
\label{P15}
\end{eqnarray}}
It is easy to check that
$\kappa_D$  satisfies the following identity}:
\begin{eqnarray}
\left( \frac{D-2}{2} g_s \frac{\partial}{\partial g_s} - 
\sqrt{\alpha'}\frac{\partial}{\partial \sqrt{\alpha'}} - \sum_{m=1}^{26-D}R_m \frac{\partial}{\partial R_m} \right) \kappa_D =0~.
\label{P7}
\end{eqnarray}
The $R_m$ dependence can also be absorbed by introducing
the $D$-dimensional string tension {$g_{D}$\cite{Ortin}}:
\begin{eqnarray}
 g_{D} \equiv {\rm e}^{\phi_0} = g_s \, G_0^{-\frac{1}{4}}
\label{P16}
\end{eqnarray}}
Then Eq. (\ref{P14}) becomes:
\begin{eqnarray}
\frac{1}{2 \kappa_D^2} = \frac{(2\pi {\sqrt{\alpha'})^{26-D}}}{ 2 g_{D}^2 \hat{\kappa}^2}~~;~~
\kappa_D 
=  (2\pi)^{\frac{D-3}{2}}\, \sqrt{2^{-9}} \,g_{D} \,(\sqrt{\alpha'})^{\frac{D-2}{2}} ~.
\label{P17}
\end{eqnarray}
and thus $\kappa_D$ also satisfies
\begin{eqnarray}
\left( \frac{D-2}{2} g_{D} \frac{\partial}{\partial g_{D}} - 
\sqrt{\alpha'}\frac{\partial}{\partial \sqrt{\alpha'}} \right) \kappa_D =0~.
\label{tree-level-soft-operator}
\end{eqnarray}

\setcounter{equation}{0}

\section{Schottky  Parametrization of Riemann Surfaces}
\label{Schottky}
 
 In this appendix we summarize the main aspects of the Schottky description of  Riemann surfaces of arbitrary genus. String amplitudes at  $h$-order in the perturbative expansion are represented by integrals over  Riemann surfaces, $\Sigma_h$, of genus $h$. The world sheet of  closed string tree diagrams, for example, is a sphere which is mapped by a stereographic projection onto $\mathbb{C}P^1$,  the complex plane with the point at infinity. This complex plane is the integration region of tree-level closed string amplitudes.
Riemann surfaces of higher order $\Sigma_h$, in the Schottky representation, are essentially identified with the complex plane  where $2h$ cycles have been cut-off and pairwise identified.

\subsection{The Schottky Group}
The basic element of the Schottky group is the projective transformation which maps 
$\mathbb{C}P^1$ to itself. It can be expressed by a matrix multiplication\cite{Gerbert}:
\begin{eqnarray}
S=\left(\begin{array}{cc} 
   a&b\\
   c&d\end{array}\right)~~;~~ S~:~\left(\begin{array}{c}
                                       z_1\\
                                       z_2\end{array}\right)\rightarrow 
                                       \left(\begin{array}{c}
                                       z'_1\\
                                       z'_2\end{array}\right) =\left(\begin{array}{cc} 
   a&b\\
   c&d\end{array}\right)\left(\begin{array}{c}
                                       z_1\\
                                       z_2\end{array}\right)
\end{eqnarray}
The entries of the matrix are complex numbers with $a\,d-b\,c=1$. The vectors\footnote{Here $t$ denotes the transpose of the vector.} $(z_1,\,z_2)^t\in \mathbb{C}^2$ are the homogeneous coordinates of  points of ${\mathbb C}P^1$. {For $z_2\neq 0$ and $z_2 ' \neq 0$  
we can define $z=z_1/z_2$ and $z' = {z_1 '}/{z_2 '}$} and  rewrite the projective transformation in the form:
\begin{eqnarray}
z'=\frac{a \,z+b}{c\,z+d} ~.\label{A.1}
\end{eqnarray}
Such a rational mapping is called  a M\"obius transformation. A property of the projective transformation is to leave invariant the cross ratio of four points:
\begin{eqnarray}
\frac{(z'_1-z'_2)(z'_3-z'_4)}{(z'_1-z'_3)(z'_2-z'_4)}= \frac{(z_1-z_2)(z_3-z_4)}{(z_1-z_3)
(z_2-z_4)}\label{A.2}
\end{eqnarray}
The fixed points $\xi$ and $\eta$ are by definition the points left invariant by the  transformation (\ref{A.1}).   These two  points together with the point $\infty$ are mapped into $\xi$, $\eta$ and $a/c$ respectively. Hence Eq.~(\ref{A.2}) for $(z_2=\eta,\,z_3=\xi,\,z_4=\infty)$ may be equivalently written    as:
\begin{eqnarray}
\frac{S(z)-\eta}{S(z)-\xi}=\mk \frac{z-\eta}{z-\xi}\qquad;\qquad 
 \mk=\frac{a-c\,\eta}{a-c\,\xi}~~;~~|\mk |\leq 1
\end{eqnarray}
where $\mk$ is the multiplier of the projective transformation $S(z)$ and it is easily seen that $\mk^n$, with $n$ a positive integer, is the multiplier of the $S^n(z)$ transformation.
It  follows that:
\begin{eqnarray}
\lim_{n \rightarrow \infty} S^n(z)=\eta\qquad,\qquad \lim_{n \rightarrow \infty} S^{-n}(z)=\xi
\end{eqnarray}
$\eta$ and $\xi$ are named attractive and repulsive fixed points, respectively. 
The infinitesimal line element $dz$ is transformed by $S(z)$ in:
\begin{eqnarray}
\frac{dz'}{dz}=S'(z)=\frac{1}{(cz+d)^2}
\end{eqnarray}
Lengths  and areas   are unaltered in magnitude if $|cz +d|=1$ For $c\neq 0$ the locus is a circle of radius $1/|c|$ and center $-d/c$. This circle is called the isometric circle ${\cal C}$. Similar considerations can be done for the inverse  transformation $S^{-1}(z)$. In this case the isometric circle, ${\cal C}'$, is $|c\,z-a|=1$ and it has radius $1/|c|$ and center $a/c$.  It is easily seen that
\begin{eqnarray}
|c\,S(z)-a|= |c\,z+d|^{-1}
\end{eqnarray}
therefore $S(z)$ maps  ${\cal C}\rightarrow {\cal C}'$  while the inverse of the transformation, $S^{-1}(z)$, transforms ${\cal C}'$ into ${\cal C}$. Moreover any point inside ${\cal C}$ is transformed by the projective transformation in a point outside the circle ${\cal C}'$ and, vice versa, the region exterior to ${\cal C}'$ is mapped by $S^{-1}$ in the interior of ${\cal C}$. In particular $\eta$ will be in the circle ${\cal C}'$ and $\xi$ in the other isometric circle\cite{DiVecchia:1988cy,Ford}.

We can now introduce the definition of Schottky group. Given a set of $h$ projective transformations $S_I$, where $I=(1\dots h)$, with the restriction that {all isometric circles are external to each other
and therefore  have zero intersection}, 
the Schottky group, {\calligra C}$_h$, is the group freely generated by the $S_I$'s~\cite{DiVecchia:1988cy,Ford}. A generic element of the group $T_\alpha\in\mbox{\calligra C}_h$, except for the identity, can be written in the form:
\begin{eqnarray}
T_\alpha= S_{I_1}^{n_1}\,S_{I_2}^{n_2}\dots S_{I_r}^{n_r},\qquad r=1,2\dots;\qquad n_i\in\mathbb{Z}/\{0\},~~I_i\neq I_{i+1}
\end{eqnarray}
The number of the generators or their inverse gives the order of the group:
\begin{eqnarray}
n_\alpha=\sum_{i=1}^r |n_r|
\end{eqnarray} 
$T_\alpha$ is a primitive element of the Schottky group if it  cannot be written as an integer power of other elements, $T_\alpha\neq (T_{\alpha'})^n$ with $n\in \mathbb{N}_+$.
The fundamental region of the group is the complex plane outside all the isometric circles $ {\cal C}_I $ and ${\cal C}'_I$ associated to each projective transformation defining the group. By identifying $ {\cal C}_I $  with ${\cal C}'_I$, $\forall I=(1\dots h)$,    
$h$ handles are formed and a Riemann surface of genus $h$ is created. More precisely:
\begin{eqnarray}
\Sigma_h=\frac{\mathbb{C} \cup \{\infty\}- \Lambda(\!\mbox{\calligra C}_h)
 }
{\mbox{\calligra C}_h}
\end{eqnarray}
 where $\Lambda(\!\mbox{\calligra C}_h)$ is the limit set of the Schottky group\cite{DiVecchia:1988cy, Bers}. 
Equivalent representations of the Schottky group are obtained by transforming each element  by a fixed projective transformation $A$, i.e. $T'_\alpha=A\,T_\alpha A^{-1}$.
Each generator of the group depends on three complex parameters, the fixed points and the multiplier, but three of them can be fixed by this overall projective transformation, therefore  inequivalent Schottky groups are parametrized by $3h-3$ complex numbers. This is exactly the number of complex moduli of a Riemann surface with  $h$ handles.

\clearpage
\subsection{The Abelian Differentials and the Period Matrix}
The canonical homology cycles $(a_I,\,b_I)$ of a Riemann surface are  identified respectively with 
${\cal C}_I$ or ${\cal C}'_I$ and with the path connecting a point $z_0\in {\cal C}_I$
with $S_I(z_0)\in {\cal C}'$. 

The  abelian differentials $\omega_I$, $I=(1\dots h)$, associated to the homology cycles are given by:
\begin{eqnarray}
\omega_I=\sum_{T_\alpha}^{(I)}\left( \frac{1}{z-T_\alpha(\eta_I)}-\frac{1}{z-T_\alpha(\xi_I)}\right)dz\equiv \omega_I(z)dz\label{abelia}
\end{eqnarray}
where $\sum_{T_\alpha}^{(I)}$ means that the sum is over all the elements of the Schottky group 
that do not have $S_I^n$, $n\in \mathbb{Z}/\{0\}$ at their  right-hand end. 
The abelian differentials are normalized in the standard way:
\begin{eqnarray}
\oint_{a_J} \omega_I=\oint_{{\cal C}'_J} \omega_I=2\pi i \delta_{IJ}\label{A.12}
\end{eqnarray} 
The normalization follows from the observation that the abelian differentials have simple poles in $T_\alpha(\eta_I)$ and $T_\alpha(\xi_I)$ and these two points are always outside the circle $a_J$ except when $S_J$ appears as leftmost factor in $T_\alpha$. In this case they  are both inside the same circle ${\cal C}'_J$ and the integral along this circle  is vanishing from the residue theorem. When $T_\alpha$ is the identity, instead, the first term of Eq.~(\ref{abelia}) gives the normalization of the integral while the second term is vanishing after integration.
The integral along the circle $b_J$ is by definition:
\begin{eqnarray}
\oint_{b_J}\omega_I=\int_{z_0}^{S_J(z_0)} \omega_I= 
\sum_{T_\alpha}^{(I)}
\log \frac{S_J(z_0)-  T_\alpha(\eta_I)}{z_0-T_\alpha(\eta_I)}\frac{z_0-T_\alpha(\xi_I)}{S_J(z_0)-T_\alpha(\xi_I)}\label{bcicle}
\end{eqnarray}
With the help of the identity
\begin{eqnarray}
\frac{(S_J(z_0)-  T_\alpha(\eta_I))(z_0-T_\alpha(\xi_I))}{(S_J(z_0)-T_\alpha(\xi_I))(z_0-T_\alpha(\eta_I))}
= \frac{(T_\alpha^{-1}S_J(z_0)-  \eta_I)(T_\alpha^{-1}(z_0)-\xi_I)}{(T_\alpha^{-1}S_J(z_0)-\xi_I)(T_\alpha^{-1}(z_0)-\eta_I)}\label{A.13}
\end{eqnarray}
the right side of the Eq.~(\ref{bcicle}) for $I\neq J$ becomes:
\ea{
\sum_{T_\alpha}^{(I)}\log \frac{(T_\alpha^{-1}S_J(z_0)-  \eta_I)(T_\alpha^{-1}(z_0)-\xi_I)}{(T_\alpha^{-1}S_J(z_0)-\xi_I)(T_\alpha(z_0)-\eta_I)}
=\sum_{T_\alpha}^{(J,\,I)}\sum_{n\in \mathbb{Z}}\log\frac{(T_\alpha^{-1}S^{n+1}_J(z_0)-  \eta_I)(T_\alpha^{-1}S^n_J(z_0)-\xi_I)}{(T_\alpha^{-1}S^{n+1}_J(z_0)-\xi_I)(T_\alpha S^n_J(z_0)-\eta_I)}
\label{A.15}
}
where $(J,\,I)$ means a sum over the elements $T_\alpha$ without $I$ and $J$ as rightmost and leftmost factor, respectively. 
It is easy to see that the sum over $n$ vanishes due to the properties of the logarithm, except at the points $n= \pm \infty$, thus
getting for $I\neq  J$
\cite{Mandelstam:1985ww}:
\eas{
\oint_{b_J}
\omega_I&=\sum_{T_\alpha}^{(J,\,I)}\log\frac{(T_\alpha^{-1}S^{\infty}_J(z_0)-  \eta_I)}{(T_\alpha^{-1}S^{\infty}_J(z_0)-\xi_I)}\frac{(T_\alpha^{-1}S^{-\infty}_J(z_0)-  \xi_I)}{(T_\alpha^{-1}S^{-\infty}_J(z_0)-\eta_I)}
\\
&=\sum_{T_\alpha}^{(J,\,I)}\log\frac{(\eta_J-  T_\alpha(\eta_I))}{(\eta_J-T_\alpha(\xi_I))}\frac{(\xi_J- T_\alpha (\xi_I))}{(\xi_J-T_\alpha(\eta_I))}	\, ,
}
where $S^\infty_J(z_0)=\eta_J$ and $S^{-\infty}(z_0)_J=\xi_J$. The same analysis can be repeated in the case $I=J$, the only difference  is that in Eq.~(\ref{A.15}) for $n\neq 0$ the  case $T_\alpha=\mathbb{I}$ has to be excluded because it violates the restriction that $T_\alpha$ doesn't have $S_I$ as rightmost factor. This means that, in using  Eq.~(\ref{A.13}), the identity has to be excluded, getting:
\ea{
{\sum}_{T_\alpha\neq \mathbb{I}}^{(I)}
\log \frac{S_I(z_0)-  T_\alpha(\eta_I)}{z_0-T_\alpha(\eta_I)}\frac{z_0-T_\alpha(\xi_I)}{S_I(z_0)-T_\alpha(\xi_I)}={\sum}_{T_\alpha\neq \mathbb{I}}^{(I,\,I)}\log\frac{(\eta_I-  T_\alpha(\eta_I))}{(\eta_I-T_\alpha(\xi_I))}\frac{(\xi_I- T_\alpha (\xi_I))}{(\xi_I-T_\alpha(\eta_I))}
}
The contribution from the identity  is instead given by:
\begin{eqnarray}
\log \frac{S_I(z_0)-  \eta_I}{S_I(z_0)-\xi_I}\frac{z_0-\xi_I}{z_0-\eta_I}=\log \mk_I
\end{eqnarray}
In conclusion we have:
\begin{eqnarray}
&&\oint_{b_J}
\omega_I=2\pi i\tau_{IJ}\label{A.19}
\end{eqnarray}
where
\begin{eqnarray}
2\pi i\tau_{IJ}= \delta_{IJ}\log \mk_I-{\sum'}_{T_\alpha}^{(J,\,I)}\log\frac{(\eta_J-T_\alpha(\xi_I))}{(\eta_J-  T_\alpha(\eta_I))}\frac{(\xi_J-T_\alpha(\eta_I))}{(\xi_J- T_\alpha (\xi_I))}
\end{eqnarray}
 and the prime on the sum means that the identity is not present for $I=J$. The quantity $\tau_{IJ}$ is the period matrix  in the Schottky representation of the Riemann surface.
 
 \subsection{The Prime Form}
The prime form $\mathbb{E}(z_1,\,z_2)$ is a differential form with conformal weight $(-1/2,\,-1/2)$. In the Schottky parametrization it is given by:
\begin{eqnarray}
{\mathbb E}(z_1,\,z_2)=\frac{(z_1-z_2)}{\sqrt{dz_1 dz_2}} \prod'_{\alpha} 
\frac{(z_1-T_\alpha(z_2))(z_2-T_\alpha(z_1))}{(z_1-T_\alpha(z_1))(z_2-T_\alpha(z_2))}
\equiv \frac{E(z_1,\,z_2)}{\sqrt{dz_1\,dz_2}}
\end{eqnarray}
where the prime  means
that the identity is excluded and $T_\alpha$ and $T_\alpha^{-1}$ are counted only once.
The prime form doesn't change if we move the  argument around the $a_I$ cycles. It changes around the $b_I$ cycles according to the rule\cite{DiVecchia:1988cy}:
\begin{eqnarray}
{\mathbb E}(S_I(z_1),\,z_2)=-e^{-i\pi\tau_{II}} \,e^{-\int_{z_2}^{z_1}\omega_I}\,{\mathbb E}(z_1,\,z_2)\label{B.176}
\end{eqnarray} 
In what follows 
we also give the periodicity properties of the function $E(z_1,z_2)$ around the 
homology cycles of the Riemann surface. This function is, by its definition,  invariant around the cycle $a_I$, while from the identity:
\begin{eqnarray}
dS_I(z)= \frac{dz}{(c_I z +d_I)^2}~~
\end{eqnarray}
we have
\ea{
{E}(S_I(z_1),\,z_2)= {\mathbb E}(S_I(z_1),\,z_2)\sqrt{dS_I(z_1)\,dz_2}= - \frac{1}{ (c_I z_1+d_I)} e^{-i\pi\tau_{II}} \,e^{-\int_{z_2}^{z_1}\omega_I}\,{ E}(z_1,\,z_2) \label{B.191}
}
In this paper, with an abuse of notation, we call both $\mathbb{E}(z_1,\,z_2)$ and $E(z_1,\,z_2)$ the prime form, although the latter is not a form.

An alternative expression of the prime form  given in terms of the Riemann Theta function is\cite{TetaII}\footnote{In Ref.~\cite{TetaII} the argument of the Theta-function is defined without the factor $1/2\pi i$ in front of the integral, i.e. $\Theta[^{\vec{\delta}_1}_{\vec{\delta}_2}](\int_{z}^w\vec{\omega}|\tau)$. However there also the  abelian differentials are normalized without that factor, i.e. $\oint_{(a_I;\,b_I)}\omega^J=(\delta_{IJ};\,\tau_{IJ})$. The factor $2\pi i$ in Eq.~(\ref{primeform1}) compensates   these different normalizations.}:
\begin{eqnarray}
\mathbb{E}=\frac{\Theta[^{\vec{\delta}_1}_{\vec{\delta}_2}](\vec{\nu}_w-\vec{\nu}_z |\tau)}{\sqrt{\zeta(w)dz}\sqrt{\zeta(w)dw}}\qquad ;\qquad\vec{\nu}_w-\vec{\nu}_z =\frac{1}{2\pi i}\int_{z}^w\vec{\omega}\label{primeform1} 
\end{eqnarray} 
where $\vec{\delta}_{1,2}$ are fixed odd theta characteristics,  $\zeta(z)=\partial_{\nu^I}\Theta[^{\vec{\delta}_1}_{\vec{\delta}_2}](0|\tau)\omega^I(z)$, and the Riemann Theta function is:
\begin{eqnarray}
\Theta \left[^{\vec{\delta}_1}_{\vec{\delta}_2}\right](\vec{\nu}|\tau)=\sum_{\vec{n}\in \mathbb{Z}^h} e^{\pi i (n^I+\delta_1^I)\tau_{IJ}(n^J+\delta_1^J)} e^{2\pi i (n_I+\delta_{1I})(\nu^I+\delta_2^I)}
\end{eqnarray}
The  Riemann Theta function  satisfies the following periodicity conditions:
 \begin{eqnarray}
\Theta \left[^{\vec{\delta}_1}_{\vec{\delta}_2}\right](\vec{\nu}+\vec{p}+\tau\vec{q}|\tau)=e^{-\pi q^I\tau_{IJ}q^J-2\pi iq_I(\nu^I+\delta_2^I)} e^{2\pi i \delta_1^Ip_I}\,\Theta \left[^{\vec{\delta}_1}_{\vec{\delta}_2}\right](\vec{\nu}|\tau)\label{trtheta}
\end{eqnarray}
with $(\vec{p},\,\vec{q})\in \mathbb{Z}^{2h}$.


\section{Scalar Green function on genus $h$ Riemann surface}
\label{AppB}

In this Appendix we discuss how the function ${\cal{G}}_h (z,w)$ introduced in Eq.~(\ref{NL6}) is related to the scalar Green function on a genus $h$ Riemann surface.

In Eq.~(2.90) of Ref.~\cite{DHoker:1988pdl}  the regularized Green function is defined through the following relation:
\ea{
G_r (z_i,  z_j) = G(z_i , z_j) - \frac{1}{2} G_R (z_i,z_i)  - \frac{1}{2} G_R (z_j, z_j) = - \frac{1}{2} \log [ \rho(z_i) \rho(z_j)]
  - \log F(z_i, z_j)
\label{NL9}
}
where the Green function is defined as follows
\begin{eqnarray}
G  (z_i, z_j) = \langle x(z_i) x(z_j) \rangle
\label{NL18}
\end{eqnarray}
and $G_R$ provides the regularization at coincident points $z_i = z_j$ (see Ref.~\cite{DHoker:1988pdl} for explicit expressions).
{The metric in the conformal gauge has been chosen to be}
\ea{
ds^2  = 
\rho (z ) \, dz \,  d {\bar{z}}
\label{ConfGauge}
}
and the function $F$, defined by
\ea{
\log F(z_i,z_j) = \log | E (z_i, z_j) |^2 + \Re \left( \int_{z_j}^{z_i} \omega _I\right) 
( 2\pi \Imt)^{-1}_{IJ} \Re \left( \int_{z_i}^{z_j} \omega _J \right)  \, ,
 \label{NL13}
}
was introduced in Eqs.~(2.90) and (2.91) of Ref.~\cite{DHoker:1988pdl},
with $E$ being the prime form, $\omega_I$ the abelian differentials, and $\tau_{IJ}$ the period matrix (see Appendix~\ref{Schottky}).
\footnote{In Ref.~\cite{DHoker:1988pdl}  the abelian differentials are normalized as
$\int_{a_J} \omega_K = \delta_{JK} \, , \ \int_{b_J} \omega_K = \tau_{JK}$
 while in our case they are normalized as (see Appendix~\ref{Schottky}):
$\int_{a_J} \omega_K = 2 \pi i\delta_{JK}~~;~~ \int_{b_J} \omega_K = 2 \pi i \tau_{JK}$.
}

If the conformal factors in \Eq{NL9} are related to the choice of coordinates $V_i (z)$, entering in Eq.~(\ref{NL6}), by 
\begin{eqnarray}
\frac{1}{\rho (z_i )} =  |V_i ' (0)|^2 ~~;~~~ \frac{1}{\rho (z_j  )} = |V_j ' (0)|^2
\label{NL12}
\end{eqnarray}
we see that Eq.~(\ref{NL9}) implies that
\begin{eqnarray}
G_r (z,w) = - {\cal{G}}_h (z,w) \, .
\label{GcalG}
\end{eqnarray}
In the following subsection we will argue that this coordinate choice can be {consistently} made. In a subsequent subsection we discuss some properties of $G_r$.


\subsection{Choice of coordinates}
\label{CooChoice}

The identification in \Eq{GcalG} is found through the relation
in Eq.~(\ref{NL12}) between $\rho (z)$ and $ V_i '(0)$.
This choice of coordinates can be made if both sides of the relation transform in the same way under conformal transformations, which we will now check.

A coordinate system around a puncture is defined by the function:
\begin{eqnarray}
w_i = V^{-1}_i (z)  \longrightarrow z = V_i (w_i)
\label{NL35}
\end{eqnarray}
where $z$ is a global coordinate that gives $z_i$ for $w_i =0$:
\begin{eqnarray}
z_i = V_i (0)
\label{NL36}
\end{eqnarray}
If we use another global parameter ${\tilde{z}}$ then we have the following relation
\begin{eqnarray}
w_i = {\tilde{V}}_i^{-1} ({\tilde{z}}) \longrightarrow {\tilde{z}} = {\tilde{V}}_i ( w_i )~~~;~~~{\tilde{z}}_i = 
{\tilde{V}}_i (0)
\label{NL37}
\end{eqnarray}
Since $w_i$ is the same  in the two equations (\ref{NL35}) and (\ref{NL37}) we get:
\begin{eqnarray}
\frac{  \partial {\tilde{z}} }{ \partial z } (w_i =0)  = \frac{ \partial {\tilde{V}} (w_i ) }{
\partial V_i (w_i) }|_{w_i=0} = \frac{{\tilde{V}}_i  '(0)}{V_i ' (0)} \Longrightarrow
{\tilde{V}}_i  ' (0) = \frac{\partial {\tilde{z}}}{\partial z}|_{w_i =0} V_i  '(0) 
\label{NL38}
\end{eqnarray}
and hence $V _i ' (0)$ transforms as
\begin{eqnarray}
{\tilde{V}}_i ' (0) {\tilde{\bar{V}}}_i ' (0) = \left(\frac{\partial {\tilde{z}}}{\partial z} \frac{ \partial {\tilde{\bar{z}}}}{\partial {\bar{z}}}\right)_{w_i=0}  V_i ' (0) {\bar{V}}_i ' (0)
\label{NL39}
\end{eqnarray}
Next we consider $\rho (z)$.
Under an arbitrary reparametrization the metric transforms as follows:
\begin{eqnarray}
g_{\mu \nu} (x) \rightarrow {\tilde{g}}_{\mu \nu} ( {\tilde{x}}) = \frac{\partial x^\rho}{\partial {\tilde{x}}^\mu}  \frac{\partial x^\sigma}{\partial {\tilde{x}}^\nu} g_{\mu \nu} (x)
\label{NL33X} 
\end{eqnarray}
In the conformal gauge the previous relation becomes
\begin{eqnarray}
\rho ( z) \rightarrow {\tilde{\rho}} ( {\tilde{z}}) = \frac{\partial z}{\partial {\tilde{z}}} 
\frac{\partial {\bar{z}} }{\partial {\bar{\tilde{z}} }} \rho (z)
\label{NL34}
\end{eqnarray}
In particular, we get
\begin{eqnarray}
\rho ( z_i) \rightarrow {\tilde{\rho}} ( {\tilde{z}}_i) = \left(  \frac{\partial z}{\partial {\tilde{z}}} 
\frac{\partial {\bar{z}} }{\partial {\bar{\tilde{z}} }}  \right)_{w_i =0} \rho (z_i)
\label{NL34a}
\end{eqnarray}
Comparing Eqs.~(\ref{NL34a}) and (\ref{NL39}), we see that they are consistent with  the identification in Eq.~(\ref{NL12}). In other words the quantity:
\begin{eqnarray}
{\tilde{\rho}} ({\tilde{z}}_i ) | {\tilde{V}}_i ' (0) |^2 = \rho (z_i) | V_i ' (0)|^2
\label{NL40}
\end{eqnarray}
is invariant under the conformal coordinate transformations.


\subsection{Properties of the regularized Green function}
The regularized Green function, $G_r(z,w)$, satisfies the following equation:
\ea{
\partial_z \partial_{ {\bar{z}}} G_r (z , w ) = -
\pi \delta^{(2)} (z -w) 
+ \frac{2 \pi g_{  z {\bar{z}} } }{\int d^2 z \sqrt{g} } \, .
\label{NL10}
}
In the conformal gauge,  $d s^2 = \rho  \, dz\,  d {\bar{z}}$,  we have that
\ea{
g_{z \bar{z}} = \rho/2 \quad \text{and} \quad \sqrt{g} = \rho/2
}
It follows that
\begin{eqnarray}
\int \partial_z \partial_{ {\bar{z}}} G_r (z , w )  d^2 z =0 \, .
\label{NL11}
\end{eqnarray}
because with our conventions for $d^2 z$ we get that \mbox{$\int d^2 z \delta^{(2)}(z-w) = 2$.
}

It is easy to verify that $G_r$ is invariant under transport along the homology  cycles of the Riemann surface.
As an example we  outline the proof in the case of the $b_I$ cycles.
With the help of Eqs.~(\ref{A.12}), (\ref{A.19}), (\ref{B.191}) and (\ref{NL34}) we get:
\ea{
-G_r(S_I(z_1),\,z_2)&=\log F(S_I(z_1),\,z_2)+\frac{1}{2}\log [\rho(S_I(z_1),\,S_I(\bar{z}_1)) \rho(z_2,\,\bar{z}_2)]\nonumber\\
&=\log F(z_1,\,z_2)-\log|c_I z+d_I|^2+\frac{1}{2}\log[\rho(z_1,\,\bar{z}_1) \rho(z_2,\,\bar{z}_2)]+\log|c_I z+d_I|^2
\nonumber \\
&=-G_r(z_1,\,z_2)
}

The $F$-function defined in \Eq{NL13}, and related to $G_r$ through \Eq{NL9}, satisfies according to Ref.~\cite{DHoker:1988pdl} the following equation:
\begin{eqnarray}
\partial_z \partial_{{\bar{z}}} \log F (z  , w) = \pi \delta^{(2)} (z-w) - \frac{1}{4 \pi} \, \omega_I (z) (\Imt )^{-1}_{IJ} {\bar{\omega}}_J ({\bar{z}})
\label{NL16}
\end{eqnarray}
On the other hand,  from Eq.~(\ref{NL9}) and (\ref{NL10}) we get
\begin{eqnarray}
\partial_z \partial_{{\bar{z}}} \log F (z , w) 
&& = 
\pi \delta^{(2)} (z-w) - \frac{2 \pi\, \rho (z) }{\int d^2 z \, \rho (z)} 
- \frac{1}{2}\partial_z \partial_{{\bar{z}}}  \log \rho (z)
\label{NL19}
\end{eqnarray}
Equating (\ref{NL16}) and (\ref{NL19}) we thus find the relation
\begin{eqnarray}
 \frac{1}{4 \pi} \omega_I (z) (\Imt )^{-1}_{IJ} {\bar{\omega}}_J ({\bar{z}}) = 
\frac{2 \pi\, \rho (z) }{\int d^2 z \, \rho (z)} 
 +\frac{1}{2}\partial_z \partial_{{\bar{z}}}  \log \rho (z)
\label{NL20}
\end{eqnarray}
We would like here to check this nontrivial relation.
We can explicitly prove the integral version of this identity.
In a two-dimensional euclidean conformally flat space-time with metric
\begin{eqnarray}
g_{ab} = \left( \begin{array}{cc} 0 &  \frac{\rho}{2} \\
\frac{\rho}{2} &0 \end{array} \right) = \frac{\rho}{2} \left( \begin{array}{cc} 0 &1 \\
1 & 0 \end{array} \right) ~~~;~~~g^{ab} =  \frac{2}{\rho}  \left( \begin{array}{cc} 0 &1 \\
1 & 0 \end{array} \right)
\label{NL21}
\end{eqnarray}
the curvature scalar is related to $\rho$ as follows
\begin{eqnarray}
R = - \frac{1}{\rho} \partial_\mu \partial^\mu \log \rho= - \frac{4}{\rho}\partial_z \partial_{\bar{z}} \log \rho 
\label{NL22}
\end{eqnarray}
 where
 \begin{eqnarray}
z = z_1 + i z_2~~;~~ {\bar{z}} = z_1 - i z_2~;~\partial_z = \frac{1}{2} \left(\partial_{z_1} - i \partial_{z_2} \right)~;~\partial_{\bar{z}} = \frac{1}{2} \left( \partial_{z_1} + i \partial_{z_2} \right)
\label{NL23}
\end{eqnarray}
Now, integrating Eq.~(\ref{NL20}) over $d^2 z$ and using 
the relation in \Eq{RBI},
\begin{eqnarray}
\int d^2 z \omega_I (z) {\bar{\omega}}_J ({\bar{z}}) ={2} (2\pi)^2 (\Imt)_{IJ}
\label{NL24}
\end{eqnarray}
we get
\begin{eqnarray}
{2}\pi h
={2} \pi - \frac{1}{8} \int d^2 z  \rho R 
\label{NL25}
\end{eqnarray}
The \emph{Gauss-Bonnet theorem} for a compact Riemann surface without boundary states that
\begin{eqnarray}
\frac{1}{4 \pi} \int \sqrt{g} R = 2(1-h)
\label{NL26}
\end{eqnarray}
and since $\sqrt{g} = \rho/2$
we see that Eq.~(\ref{NL25}) is indeed satisfied. 

We can furthermore check \Eq{NL20} explicitly at one loop. 
At one-loop we have
\begin{eqnarray}
\omega (z) = \frac{1}{z}~~;~~~{\bar{\omega}} ({\bar{z}}) = \frac{1}{\bar{z}}~~;~~~ \Imt \equiv \tau_2
~~;~~ R=0
\label{NL28}
\end{eqnarray}
Then, Eq.~(\ref{NL20}), using \Eq{NL22}, becomes
\begin{eqnarray}
\frac{1}{4 \pi |z|^2 \tau_2} = \frac{{2}\pi \rho (z)}{\int d^2 z \rho (z)}
\label{NL29}
\end{eqnarray}
Assuming that, at one-loop, $\rho (z)$ is equal to (which we prove in the end)
\begin{eqnarray}
\rho (z) = \frac{1}{(2\pi)^2 |z|^2}= \frac{1}{(2\pi)^2 } \omega (z) {\bar{\omega}} ({\bar{z}})
\label{NL30}
\end{eqnarray}
we can compute the volume integral in the denominator of the rhs
\begin{eqnarray}
\int d^2 z \rho (z)&& = \frac{1}{(2\pi)^2} \int \frac{d^2 z}{|z|^2} = \frac{1}{(2\pi)^2}  \int {\frac{dz \wedge d {\bar{z}}}{i z}} \partial_{\bar{z}} \log |z|^2 \nonumber \\
&& = \frac{1}{(2\pi)^2}  \left( \oint_{|z|=1} {\frac{dz}{iz}} \log |z|^2 - \oint_{|z|=|\mk|}   {\frac{dz}{iz}} \log |z|^2 \right) = -   \frac{{2}\pi}{(2\pi)^2} \log |\mk|^2  \nonumber \\
&& = {2} \tau_2
\label{NL31}
\end{eqnarray}
where we have used the relation:
\begin{eqnarray}
|\mk|^2 = {\rm e}^{-4 \pi \tau_2} 
\label{NL32}
\end{eqnarray}
and the fact that the integral is performed in the region of the $z$-plane contained between the circle of radius $1$ and that of radius $|\mk|$. 
Consistently, the result of \Eq{NL31} follows also immediately from \Eq{NL24} and \Eq{NL30}.
Eqs.~(\ref{NL30}) and (\ref{NL31}) imply that Eq.~(\ref{NL29}) is satisfied.

Finally, comparing Eqs.~(\ref{NL12}) and (\ref{NL30}), at one-loop,  we get\cite{DiVecchia:NP96}:
\begin{eqnarray}
V_i ' (0) = 2\pi z_i
\label{NL33}
\end{eqnarray}
The same result can be obtained by observing that the one-loop world-sheet in 
{the} closed string  is a torus which is a flat manifold. Parametrizing the manifold with real coordinates $(\tau,\,\sigma)$,  the metric will be given by $ds^2=\d\tau^2+d\sigma^2$. By  rewriting the  metric in terms of the complex coordinates $z=e^{2\pi (\tau-i\sigma)}$ and $\bar{z}=e^{2\pi (\tau+i\sigma)}$, one gets:
\begin{eqnarray}
ds^2=\d\tau^2+d\sigma^2= \frac{dzd\bar{z}}{(2\pi)^2|z|^2}
\end{eqnarray}  
It is then immediately seen that the 1-loop conformal factor of the metric is exactly the one written in Eq.~(\ref{NL30}).


\section{Derivation of the $N$-Reggeon vertex}
\label{NRegge}
\setcounter{equation}{0}

In this Appendix we derive the $N$-Reggeon Vertex for the closed bosonic string given in Eq.~(\ref{NL1}) starting from Eq. 
(3.24) of Ref.~\cite{DiVecchia:1988cy}. We start by noticing that the terms in the first line of Eq.~(3.23)
of the reference cancel with the terms in the denominator in the log of the second line of Eq.~(3.24) with $i \neq j$.
This happens for both the holomorphic and anti-holomorphic variables. This means that the vertex in
Eq.~(3.24) of Ref.~\cite{DiVecchia:1988cy} can be written as follows:
\begin{eqnarray}
V_N = && C_h N_0^N
\int dV \langle \Omega | \prod_{i=1}^{N}  [ | V_i' (0) |^{\frac{\alpha' p_i^2}{2} }]
\exp \left[\frac{1}{2} \sum_{i=1}^N \sum_{n=1}^\infty \frac{\alpha_n^{(i)}}{n!} \alpha_0^{(i)} \frac{\partial^n}{\partial z^n} \log V ' (z) \Big|_{z=0}\right]
\nonumber \\
&&  \times
\exp \left[\frac{1}{2} \sum_{i=1}^N \sum_{n=1}^\infty \frac{{\bar{\alpha}}_n^{(i)}}{n!} \alpha_0^{(i)} \frac{\partial^n}{\partial {\bar{z}}^n} \log {\bar{V}} ' ({\bar{z}}) \Big|_{{\bar{z}}=0}\right]
\nonumber \\
&& \times \exp \left[  \frac{1}{2}\sum_{i \neq j} \sum_{n,m=0}^{\infty} \frac{\alpha_{n}^{(i)}}{n!}  \partial_z^n \partial_y^m \log E ( V_i (z) , V_j (y) )|_{z=y=0} \frac{\alpha_m^{(j)}}{m!} \right] \nonumber \\
&& \times \exp \left[  \frac{1}{2}\sum_{i \neq j} \sum_{n,m=0}^{\infty} \frac{{\bar{\alpha}}_{n}^{(i)}}{n!}  \partial_{\bar{z}}^n \partial_{\bar{y}}^m \log E ( {\bar{V}}_i ({\bar{z}}) , {\bar{V}}_j ({\bar{y}}) )|_{z=y=0} \frac{{\bar{\alpha}}_m^{(j)}}{m!} \right] \nonumber \\
&& \times \exp \left[ \frac{1}{2} \sum_{i=1}^N \sum_{n,m=0}^{\infty} \alpha_n^{(i)} \frac{1}{n! m!} \partial_z^n \partial_y^m \log \frac{ E (V_i (z) , V_{i} (y) )}{V_i (z) - V_i (y) }|_{z=y=0}  \alpha_m^{(i)}\right]
\nonumber \\
&& \times \exp \left[ \frac{1}{2} \sum_{i=1}^N
\sum_{n,m=0}^{\infty} {\bar{\alpha}}_n^{(i)} \frac{1}{n! m!} \partial_{\bar{z}}^n \partial_{\bar{y}}^m \log \frac{ E ({\bar{V}}_i ({\bar{z}}) , {\bar{V}}_{i} ({\bar{y}})) }{{\bar{V}}_i ({\bar{z}}) - {\bar{V}}_i ({\bar{y}}) }|_{{\bar{z}}={\bar{y}}=0}  {\bar{\alpha}}_m^{(i)}\right]
\nonumber \\
&& \times \exp \left[ \sum_{i, j=1}^{N} \sum_{n=0}^{\infty} \left( \frac{\alpha_n^{(i)}}{n!} \partial_z^n +
\frac{{\bar{\alpha}}_n^{(i)}}{n!} \partial^n_{ {\bar{z}} } \right) \Re \left( \int_{z_0}^{V_i (z)} \omega_I 
\right)  (2\pi \Imt)^{-1}_{IJ}  \right. \nonumber \\
&&  \times \left.  \sum_{m=0}^{\infty} \left( \frac{\alpha_m^{(j)}}{n!} \partial_z^m +
\frac{{\bar{\alpha}}_m^{(i)}}{m!} \partial^m_{ {\bar{z}} } \right) \Re \left( \int_{z_0}^{V_j (z)} \omega_J 
\right)  \right]
\label{NL1a}
\end{eqnarray}
where we have also integrated over the momenta circulating in the $h$ loops.  The final vertex is written in Eq.~(\ref{NL1}) where, for reasons that will become clear later, using momentum conservation,  we have put the factor in front containing $V_i ' (0)$ together with the prime-form.
This has been done by rewriting it as follows:
\begin{eqnarray}
\prod_{i=1}^{N}  [  (V_i' (0))^{\frac{\alpha' p_i^2}{2} }] = \exp \left[ - \frac{\alpha'}{2} \sum_{i<j=1}^Np_ip_j\log 
\sqrt{V_i ' (0) V_j ' (0)}  \right]
\label{NLxx}
\end{eqnarray} Re
In the vertex in Eq.~(\ref{NL1}) we have also eliminated the dependence on the arbitrary point $z_0$.
This is shown in the following.

Let us consider the part containing the momentum given by
\begin{eqnarray}
&&\sum_{i,j} p_i p_j  \Re \left(\int_{z_0}^{V_i (z)} \omega_I \right)
 (2\pi \Imt)^{-1}_{IJ} \Re \left(\int_{z_0}^{V_j (y)} \omega_J \right)|_{z=y=0}\nonumber \\
&& = \sum_{i \neq j} p_i p_j  \Re \left(\int_{z_0}^{V_j (y)}  + \int_{V_j (y)}^{V_i (z)} \right) \omega_I 
(2\pi \Imt)^{-1}_{IJ} \Re \int_{z_0}^{V_j (y)} \omega_J|_{z=y=0}  \nonumber \\
&&\quad + \sum_{i} p_i^2 \Re \left(\int_{z_0}^{V_i (z)} \omega_I \right) (2\pi \Imt)^{-1}_{IJ} \Re \left(\int_{z_0}^{V_i (z)} \omega_J \right)|_{z=y=0}  \nonumber \\
&& = \sum_{i \neq j} p_i p_j  \Re \left( \int_{V_j (y)}^{V_i (z)}  \omega_I  \right) 
(2\pi \Imt)^{-1}_{IJ} \Re \left(\int_{z_0}^{V_j (y)} \omega_J \right)\Bigg{|}_{z=y=0}  \nonumber \\
&&\quad - \sum_j p_j^2  \Re \left( \int_{z_0}^{V_j (y)}  \omega_I \right) (2\pi \Imt)^{-1}_{IJ} \Re \left(\int_{z_0}^{V_j (y)} \omega_J \right) \Bigg{|}_{z=y=0}   \nonumber \\
&& \quad+ \sum_{i} p_i^2 \Re \left(\int_{z_0}^{V_i (z)} \omega_I \right) (2\pi \Imt)^{-1}_{IJ} \Re \left(\int_{z_0}^{V_i (z)} \omega_J \right)|_{z=y=0}  \nonumber \\
&& = \sum_{i \neq j} p_i p_j  \Re \left( \int_{V_j (y)}^{V_i (z)}  \omega_I  \right) 
(2\pi \Imt)^{-1}_{IJ} \Re \left(\int_{z_0}^{V_j (y)} \omega_J \right)\Bigg{|}_{z=y=0} 
\label{NL6a}
\end{eqnarray}
In conclusion, we get 
\begin{eqnarray}
&&\sum_{i,j} p_i p_j  \Re \left( \int_{z_0}^{V_i (z)} \omega_I \right)
(2\pi \Imt)^{-1}_{IJ} \Re \left(\int_{z_0}^{V_j (y)} \omega_J \right) \Bigg{|}_{z=y=0}\nonumber \\
&& =  \sum_{i , j} p_i p_j  \Re \left( \int_{V_j (y)}^{V_i (z)}  \omega_I  \right) 
(2\pi \Imt)^{-1}_{IJ} \Re \left(\int_{z_0}^{V_j (y)} \omega_J \right)\Bigg{|}_{z=y=0} 
\label{NL6b}
\end{eqnarray}
where in the last step we have added the terms $i=j$ because they do not give any contribution.
In order to see that the quantity on the lhs of the previous equation {is independent of $z_0$}, we rewrite it as follows:
\begin{eqnarray}
&&\sum_{i,j} p_i p_j  \Re \left( \int_{z_0}^{V_i (z)} \omega_I \right)
(2\pi \Imt)^{-1}_{IJ} \Re \left(\int_{z_0}^{V_j (y)} \omega_J \right) \Bigg{|}_{z=y=0}\nonumber \\
&& = \sum_{i,j} p_i p_j  \Re \left[ \left( \int_{z_0}^{V_j (y)} + \int_{V_j (y)}^{V_i (z)} \right)  \omega_I \right]
(2\pi \Imt)^{-1}_{IJ}  
 \Re \left[ \left(\int_{z_0}^{V_i (z)} + \int_{V_i (z)}^{V_j (y)} \right) \omega_J \right] \Bigg{|}_{z=y=0}\nonumber \\
\label{NL6c}
\end{eqnarray}
that implies
\begin{eqnarray}
&& \sum_{i,j} p_i p_j  \Re  \left(  \int_{V_j (y)}^{V_i (z)} \right)  \omega_I 
(2\pi \Imt)^{-1}_{IJ}   \Re  \left( \int_{V_i (z)}^{V_j (y)} \right) \omega_J  \Bigg{|}_{z=y=0}
\nonumber \\
&& = - \sum_{i,j} p_i p_j  \Re  \left( \int_{z_0}^{V_j (y)}  \right)  \omega_I 
(2\pi \Imt)^{-1}_{IJ} \Re  \left( \int_{V_i (z)}^{V_j (y)} \right) \omega_J  \Bigg{|}_{z=y=0}\nonumber \\
&& \quad- \sum_{i,j} p_i p_j  \Re  \left(  \int_{V_j (y)}^{V_i (z)} \right)  \omega_I  (2\pi \Imt)^{-1}_{IJ} 
\Re  \left(\int_{z_0}^{V_i (z)}  \right) \omega_J  \Bigg{|}_{z=y=0} \nonumber \\
&& = -2 \sum_{i,j} p_i p_j  \Re  \left(  \int_{V_i (z)}^{V_j (y)} \right)  \omega_I  (2\pi \Imt)^{-1}_{IJ} 
\Re  \left(\int_{z_0}^{V_j (y)}  \right) \omega_J  \Bigg{|}_{z=y=0}
\label{NL6d}
\end{eqnarray}
where we have used the fact that the terms in the second and third lines are equal as one can see  by exchanging the indices  $i$ with $j$ in one of them.  Comparing  the previous equation with Eq.~(\ref{NL6b}) we get the final result
\begin{eqnarray}
&&\sum_{i,j} p_i p_j  \Re \left( \int_{z_0}^{V_i (z)} \omega_I \right)
(2\pi \Imt)^{-1}_{IJ} \Re \left(\int_{z_0}^{V_j (y)} \omega_J \right) \Bigg{|}_{z=y=0}\nonumber \\
&& = \frac{1}{2} \sum_{i,j} p_i p_j  \Re  \left(  \int_{V_j (y)}^{V_i (z)}  \omega_I \right) 
(2\pi \Imt)^{-1}_{IJ}   \Re  \left( \int_{V_i (z)}^{V_j (y)}  \omega_J \right)  \Bigg{|}_{z=y=0}
\label{AppFin}
\end{eqnarray}
that shows independence on $z_0$ for the terms with two momenta. The terms involving only the oscillators do not have any dependence on $z_0$.  The terms with an oscillator and a momentum
do not depend on $z_0$ because of momentum conservation. Therefore we can eliminate $z_0$ everywhere as in Eq.~(\ref{AppFin}). 
This ends the derivation of Eq.~(\ref{NL1}).



\section{Calculation of $S_1$}
\label{CalS1}
\setcounter{equation}{0}

Let us for brevity make some definitions and recall some identities,
\begin{align}
&\partial_{\bar{z}_1}\partial_{z_1}{\cal G}(z_1,\,z_2) = \pi \delta^{(2)}(z_1-z_2)+ {\cal T} (z_1)
\label{boxG}
\\[2mm]
&\int_{\Sigma_h}d^2z \partial_{\bar{z}}\partial_{z}{\cal G}(z,\,w)=0
\quad 
\Leftrightarrow \quad \int_{\Sigma_h}d^2z {\cal T}(z) = -{2} \pi
\label{intT}
\\[2mm]
& e^{\frac{\alpha'}{2} k_i q {\cal G}(z_i, z_i) }  \propto | E(z_i,z_i ) |^{\alpha' k_i q } = 0
\label{eGzero}
\end{align}
where $\G$ is the quantity defined in \Eq{NL6}, however, in this section we suppress the index $h$ for brevity.
These relations apply for any number of loops $h \geq 1$, and are sufficient to determine the soft behavior of $S_1$ defined in \Eq{S1} through order $q$ in terms of $\cal G$ only, which will here be shown. 
We repeat the definition of $S_1$ and decompose the integrand in three parts:
\ea{
S_1 = &
\int \d^2 z  \, 
\left [ 1 + \sum_{j\neq i}^n  \frac{\a'}{2} k_j q  \mathcal{G}_h(z_i, z) + \frac{1}{2} \left (\frac{\a'}{2}\right )^2 
\sum_{j, l\neq i}^n  (k_j q) (k_l q)  \mathcal{G}_h(z_j, z) \mathcal{G}_h(z_l, z) \right ]
\nonumber \\
&\qquad \times
\frac{\a'}{2} \sum_{i,j=1}^n   (k_i \e) (k_j \bar{\e}) \partial_{z} \mathcal{G}_h(z_i,z)\partial_{\bar{z}} \mathcal{G}_h(z_j,z)e^{ \frac{\a'}{2} k_i q  \mathcal{G}_h(z_i, z) }
+ \Ord(q^2) 
\nonumber \\
=&  I_0 + \left (\frac{\a'}{2}\right ) I_1 + \frac{1}{2}\left (\frac{\a'}{2}\right )^2 I_2 + \Ord(q^2)
\label{App:S1}
}

In the following subsections we will make substantial use of {integration by parts}, neglecting {boundary} (total derivative) terms, to compute the three terms above. Remarkably, the three terms can be fully computed by only using the above properties of Green's function; i.e. without knowing its explicit form, in the case where the soft state is symmetrically polarized. We summarize here the results of the calculation:
\ea{
I_0 =&{2}\pi \sum_i\frac{(k_i\epsilon_q)(k_i\bar{\epsilon}_q)}{k_iq}-2\pi \frac{\a'}{2}
\sum_{i\neq j} (k_i\epsilon_q)(k_j\bar{\epsilon}_q) {\cal G}(z_j,\,z_i)\Big[1+\frac{\alpha'}{4}k_iq {\cal G}(z_j,\,z_i)\Big] \\[2mm]
I_1 =&{2}
\pi \epsilon_{q\mu\nu}^S \sum_{i\neq j} \frac{k_i^\mu k_i^\nu}{k_iq} (k_jq)  {\cal G}(z_j,\,z_i)
+ \frac{{2}\pi \alpha'}{4}  \epsilon_{q\mu\nu}^S \sum_{i\neq j } k_i^\mu k_j^\nu (k_jq) {\cal G}^2(z_i,\,z_j)
\nonumber\\
&
+ \frac{{2}\pi \alpha'}{4}  \epsilon_{q\mu\nu}^S \sum_{i\neq j , l} 
 \left [k_j^\mu k_l^\nu (k_i q) -  k_i^\mu k_l^\nu (k_jq)- k_i^\mu k_j^\nu (k_l q)  \right ]{\cal G}(z_i,\,z_j) {\cal G}(z_i,\,z_l) \\[2mm]
I_2 
=&{2}\pi \sum_{i=1}(k_i\epsilon_q)(k_i\bar{\epsilon}_q) \sum_{j,l\neq i} \frac{ (k_jq)(k_lq)    }{qk_i}{\cal G}(z_i,\,z_j)\,{\cal G}(z_i,\,z_l)
}
Only in computing a particular integral in $I_1$ was it necessary to impose polarization symmetry on $\epsilon_{q\mu \nu} = \e_{q\mu} \bar{\e}_{q\nu}$, and this was in fact not necessary for the leading order $\Ord(q^0)$ term in $I_1$. Thus through order $q^0$ the result is valid for any of the three physical states of the massless closed string; the graviton, the dilaton and the Kalb-Ramond B-field.

\subsection{Calculation of $I_0$}
\label{I0}

The integral $I_0$ can be written as:
\ea{
 I_0 
&= \sum_{i,j} \frac{(k_i\epsilon_q)(k_j\bar{\epsilon}_q)}{k_i q}\int d^2 z  \left [ \partial_{{z}}\Big(\partial_{\bar{z}}  {\cal G}(z,\,z_j)~ e^{\frac{\alpha'}{2} k_iq {\cal G}(z,\,z_i)}\Big)
-
\partial_{\bar{z}}\partial_z  {\cal G}(z,\,z_j)~ e^{\frac{\alpha'}{2} k_iq {\cal G}(z,\,z_i)}
\right ]
}
The boundary term vanishes, and in the second term the definition in \Eq{boxG} can be inserted to get
\ea{
I_0 &=-\sum_{i,j} \frac{(k_i\epsilon_q)(k_j\bar{\epsilon}_q)}{k_i q}\int d^2 z  \big[\pi \delta^2(z-z_j)+{\cal T}(z) \Big]~ e^{\frac{\alpha'}{2} k_iq {\cal G}(z,\,z_i)}\nonumber\\
&=-{2}\pi\sum_{i,j} \frac{(k_i\epsilon_q)(k_j\bar{\epsilon}_q)}{k_i q}
e^{\frac{\alpha'}{2} k_iq {\cal G}(z_j,\,z_i)}
-\sum_{i} \frac{(k_i\epsilon_q)(- q \bar{\epsilon}_q)}{k_i q}
\int d^2 z ~ {\cal T}(z) e^{\frac{\alpha'}{2}k_iq{\cal G}(z,\,z_i)}
\nonumber\\
&=-{2}\pi\sum_{i\neq j} \frac{(k_i\epsilon_q)(k_j\bar{\epsilon}_q)}{k_i q}
e^{\frac{\alpha'}{2} k_iq {\cal G}(z_j,\,z_i)}
}
To arrive at the last equality we used identity~\eqref{eGzero} as well as $\sum_j (k_j \bar{\epsilon}_q ) = - ( q \bar{\epsilon}_q ) = 0$. 

By expanding the previous expression in the soft-momentum, the first term of the expansion is the Weinberg soft theorem, i.e.:
\ea{
\sum_{i\neq j} \frac{(k_i\epsilon_q)(k_j\bar{\epsilon}_q)}{k_iq}=-\sum_{ i}
 \frac{(k_i\epsilon_q)((k_i+q)\bar{\epsilon}_q)}{k_iq}=-\sum_i\frac{(k_i\epsilon_q)(k_i\bar{\epsilon}_q)}{k_iq}
 }
 Thus, in conclusion:
\ea{
I_0 &={2}\pi \sum_i\frac{(k_i\epsilon_q)(k_i\bar{\epsilon}_q)}{k_iq}-{2}\pi \frac{\a'}{2}
\sum_{i\neq j} (k_i\epsilon_q)(k_j\bar{\epsilon}_q) {\cal G}(z_j,\,z_i)\Big[1+\frac{\alpha'}{4}k_iq {\cal G}(z_j,\,z_i)\Big]+O(q^2) 
}
This result is formally identically to the tree-level result in Ref.~\cite{DiVecchia:2015oba}, with the difference being in the Green function.

\subsection{Calculation of $I_1$}
The integral $I_1$ can be written as
\ea{
I_1 
=&\sum_{ij} \frac{(k_i\epsilon_q)(k_j\bar{\epsilon}_q)}{k_iq} \sum_{l\neq i}  (k_lq) \int d^2z\, {\cal G}(z_,\,z_l) \partial_{\bar{z}}{\cal G}(z,\,z_j)~ \partial_ze^{\frac{\alpha'}{2} k_iq {\cal G}(z,\,z_i)}
}
and integration by parts gives:
\ea{
I_1
=&
-\sum_{ij} \frac{(k_i\epsilon_q)(k_j\bar{\epsilon}_q)}{k_iq} \sum_{l\neq i} (k_lq) \int d^2z\,\partial_{\bar{z}}{\cal G}(z_,\,z_j) \partial_z{\cal G}(z,\,z_l)  e^{\frac{\alpha'}{2} k_iq {\cal G}(z,\,z_i)}\nonumber\\
&-\sum_{ij} \frac{(k_i\epsilon_q)(k_j\bar{\epsilon}_q)}{k_iq} \sum_{l\neq i}  (k_lq) \int d^2z\,{\cal G}(z_,\,z_l) \partial_{\bar{z}}\partial_z{\cal G}(z,\,z_j)  e^{\frac{\alpha'}{2} k_iq {\cal G}(z,\,z_i)} 
\label{nextI-2}
}
The integral in the second line gives:
\ea{
I_{1}^{(2)} = 
-\sum_{ij} \frac{(k_i\epsilon_q)(k_j\bar{\epsilon}_q)}{k_iq} \sum_{l\neq i}  (k_lq)\int d^2z\,{\cal G}(z,\,z_l) \Big[\pi \delta^2(z-z_j)+{\cal T}(z)  \Big]e^{\frac{\alpha'}{2} k_iq {\cal G}(z,\,z_i)} 
}
Notice that for the second term we can replace by momentum conservation
$\sum_jk_j\bar{\epsilon}_q=-q\epsilon_q=0$, and hence only the first term remains, which after expansion reads:
\ea{
I_{1}^{(2)}
&=-{2}\pi\sum_{i\neq j} \frac{(k_i\epsilon_q)(k_j\bar{\epsilon}_q)}{k_iq} \sum_{{\color{red}l\neq i}}  (k_lq) \,{\cal G}(z_j,\,z_l)\Big[1+\frac{\alpha'}{2} k_iq {\cal G}(z_j,\,z_i)\Big]+O(q^2)\label{3.101}
}
where we used that for $i=j$ the expression is zero, before expansion, due to \Eq{eGzero}. 
For {\color{red} $l=j$} there is a divergence, but as we will see, it cancels against another term, coming from the first integral in \Eq{nextI-2}.

To calculate the first integral in \Eq{nextI-2}, we consider the cases $i=j$ and $i\neq j$ separately.
The case $i=j$ can be written as:
\ea{
I_1^{(1)} \big |_{i=j} &= -\sum_{i} \frac{(k_i\epsilon_q)(k_i\bar{\epsilon}_q)}{k_iq} 
\sum_{l\neq i} \frac{ (k_lq)}{\frac{\alpha'}{2}(k_iq)} \int d^2z\,\partial_{{z}}{\cal G}(z_,\,z_l) \partial_{\bar{z}}  e^{\frac{\alpha'}{2} k_iq {\cal G}(z,\,z_i)}\nonumber\\
&=
+\sum_{i} \frac{(k_i\epsilon_q)(k_i\bar{\epsilon}_q)}{k_iq} \sum_{l\neq i} \frac{ (k_lq)}{\frac{\alpha'}{2}(k_iq)} \int d^2z\,\partial_z \partial_{\bar{z}}{\cal G}(z_,\,z_l)  e^{\frac{\alpha'}{2} k_iq {\cal G}(z,\,z_i)}
}
Using \Eq{intT} and \Eq{boxG} we get:
\ea{
I_1^{(1)} \big |_{i=j}
=&{2}\pi\sum_{i} \frac{(k_i\epsilon_q)(k_i\bar{\epsilon}_q)}{k_iq} \sum_{l\neq i} \frac{ (k_lq)}{\frac{\alpha'}{2}(k_iq)} \Big[ e^{\frac{\alpha'}{2} k_iq {\cal G}(z_l,\,z_i)}-1\Big]\nonumber\\
&+\sum_{i} \frac{(k_i\epsilon_q)(k_i\bar{\epsilon}_q)}{k_iq}\sum_{l\neq i}(k_lq)  \int d^2z {\cal T}(z) {\cal G}(z,\,z_i)\nonumber\\
&+\frac{\alpha'}{4}\sum_{i} (k_i\epsilon_q)(k_i\bar{\epsilon}_q) \sum_{l\neq i} (k_lq)  \int{d^2z}{\cal T}(z) {\cal G}^2(z,\,z_i)+O(q^2)\nonumber\\
=&{2}\pi\sum_{i} \frac{(k_i\epsilon_q)(k_i\bar{\epsilon}_q)}{k_iq} \sum_{l\neq i} (k_lq)  {\cal G}(z_l,\,z_i)+\frac{{2}\pi\alpha'}{4}\sum_{i} (k_i\epsilon_q)(k_i\bar{\epsilon}_q) \sum_{l\neq i} (k_lq) {\cal G}^2(z_l,\,z_i)\nonumber\\
&{\color{blue}-\sum_{i}(k_i\epsilon_q)(k_i\bar{\epsilon}_q) \int d^2z {\cal T}(z) {\cal G}(z,\,z_i)}\nonumber\\
&-\frac{\alpha'}{4}\sum_{i}(k_i\epsilon_q)(k_i\bar{\epsilon}_q)(k_iq) \int{d^2z}{\cal T}(z) {\cal G}^2(z,\,z_i)+O(q^2)\label{1.45-2}
}
where  we have used $\sum_{l\neq i}k_lq=-k_iq$ in the last two terms. We will see in a moment that the second-to-last ({\color{blue} blue}) line cancels out with another term.

In the case $i\neq j$ we make instead the following rewriting:
\ea{
I_1^{(1)} \big |_{i\neq j} 
=&-\sum_{i\neq j} \frac{(k_i\epsilon_q)(k_j\bar{\epsilon}_q)}{k_iq}\sum_{l\neq i} (k_lq)\int d^2z\, \partial_{\bar{z}}\Big[\partial_{{z}}{\cal G}(z,\,z_l) {\cal G}(z,\,z_j)\Big] e^{\frac{\alpha'}{2}k_iq {\cal G}(z,\,z_i)}\nonumber\\
&+\sum_{i\neq j} \frac{(k_i\epsilon_q)(k_j\bar{\epsilon}_q)}{k_iq}\sum_{l\neq i} (k_lq) \int d^2z\, 
\partial_z \partial_{\bar{z}}{\cal G}(z,z_l)
 {\cal G}(z,\,z_j) e^{\frac{\alpha'}{2}k_iq {\cal G}(z,\,z_i)}
 }
 We can expand the exponentials, since around the poles of the integrand, $z=z_l$ and $z=z_j$, the exponential is  regular ($i\neq j, l$). This gets rid of a total derivative in the first term, and by using \Eq{boxG} in the second term one finds:
 \ea{
 I_1^{(1)} \big |_{i\neq j} 
=&-\frac{\alpha'}{2}\sum_{i\neq j} (k_i\epsilon_q)(k_j\bar{\epsilon}_q)\sum_{l\neq i} (k_lq) \int d^2z\, \partial_{\bar{z}}\Big[ \partial_{{z}}{\cal G}(z_,\,z_l) {\cal G}(z,\,z_j)\Big]{\cal G}(z,\,z_i)\nonumber\\
&{\color{red}+{2}\pi\sum_{i\neq j} \frac{(k_i\epsilon_q)(k_j\bar{\epsilon}_q)}{k_iq}\sum_{l\neq i} (k_lq) {\cal G}(z_l,\,z_j)\Big[1+ {\frac{\alpha'}{2}k_iq {\cal G}(z_l,\,z_i)}\Big]}\nonumber\\
&{\color{blue}-\sum_{i\neq j} (k_i\epsilon_q)(k_j\bar{\epsilon}_q) \int {d^2z}{\cal T}(z)\,  {\cal G}(z,\,z_j)\Big[1}+\frac{\alpha'}{2}k_iq {\cal G}(z,\,z_i)\Big] +O(q^2)\label{1.46-2}
}
By summing the ({\color{red} red}) second term with Eq.~(\ref{3.101}) we get:
\ea{
&
{\color{red}-{2}\pi\sum_{i\neq j} \frac{(k_i\epsilon_q)(k_j\bar{\epsilon}_q)}{k_iq} \sum_{l\neq i}  (k_lq) \,{\cal G}(z_j,\,z_l)\Big[1+\frac{\alpha'}{2} k_iq {\cal G}(z_j,\,z_i)\Big]}\nonumber\\
&{\color{red} +{2}\pi\sum_{i\neq j} \frac{(k_i\epsilon_q)(k_j\bar{\epsilon}_q)}{k_iq}\sum_{l\neq i} (k_lq) {\cal G}(z_l,\,z_j)\Big[1+ {\frac{\alpha'}{2}k_iq {\cal G}(z_l,\,z_i)}\Big]}\nonumber\\
&=\frac{{2}\pi \alpha' }{2}\sum_{i\neq j} (k_i\epsilon_q)(k_j\bar{\epsilon}_q)\sum_{l\neq i} (k_lq) {\cal G}(z_l,\,z_j)\Big[ {\cal G}(z_l,\,z_i)- {\cal G}(z_j,\,z_i)\Big]
}
This removes, as promised, the divergent terms for $l = j$. In the last line $l=j$ is evidently zero, and the sum over $l$ can be reduces to $l\neq i,j$.

By summing the ({\color{blue} blue}) first term in the last line of Eq.~(\ref{1.46-2}) with the second-to-last line in Eq.~(\ref{1.45-2}) ({\color{blue} also blue}), we get:
\ea{
&{\color{blue}-\sum_{i}(k_i\epsilon_q)(k_i\bar{\epsilon}_q)   \int{d^2z}{\cal T}(z) {\cal G}(z,\,z_i)-\sum_{i\neq j} (k_i\epsilon_q)(k_j\bar{\epsilon}_q) \int {d^2z}{\cal T}(z)\,  {\cal G}(z,\,z_j)}
\nonumber\\
&{\color{blue} =-\sum_{i, j} (k_i\epsilon_q)(k_j\bar{\epsilon}_q)  \int {d^2z}{\cal T}(z)\,  {\cal G}(z,\,z_j)=\sum_{ j} (q\epsilon_q)(k_j\bar{\epsilon}_q)  \int {d^2z}{\cal T}(z)\,  {\cal G}(z,\,z_j)} =0
}
By collecting the results for all $i$ and $j$ we get:
\ea{
I_1 =&{2}
\pi\sum_{i} \frac{(k_i\epsilon_q)(k_i\bar{\epsilon}_q)}{k_iq} \sum_{l\neq i} (k_lq)  {\cal G}(z_l,\,z_i)
+\frac{{2}\pi\alpha'}{4}\sum_{i} (k_i\epsilon_q)(k_i\bar{\epsilon}_q) \sum_{l\neq i} (k_lq) {\cal G}^2(z_l,\,z_i)\nonumber\\
& +
\frac{{2}\pi\alpha'}{2}\sum_{i\neq j} (k_i\epsilon_q)(k_j\bar{\epsilon}_q)\sum_{l\neq i} (k_lq) {\cal G}(z_l,\,z_j)\Big[ {\cal G}(z_l,\,z_i)- {\cal G}(z_j,\,z_i)\Big]\nonumber\\
&-\frac{\alpha'}{2}\sum_{i\neq j} (k_i\epsilon_q)(k_j\bar{\epsilon}_q)\sum_{l\neq i} (k_lq) \int d^2z\, \partial_{\bar{z}}\Big[ \partial_{{z}}{\cal G}(z_,\,z_l) {\cal G}(z,\,z_j)\Big]{\cal G}(z,\,z_i)\nonumber\\
&{\color{purple}-\frac{\alpha'}{4}\sum_{i}(k_i\epsilon_q)(k_i\bar{\epsilon}_q)(k_iq)  \int{d^2z}{\cal T}(z) {\cal G}^2(z,\,z_i)}\nonumber\\
& {\color{orange}-\frac{\alpha'}{2}\sum_{i\neq j} (k_i\epsilon_q)(k_j\bar{\epsilon}_q)k_iq  \int {d^2z}{\cal T}(z)\,  {\cal G}(z,\,z_j) {\cal G}(z,\,z_i)} +O(q^2)\label{1.65-2}
}
The first term is the only one of order $q^0$ and it is formally identical to the tree-level results.
There are still three integrals to be computed at the order $q$. We will show that the first of these, not involving $\mathcal{T}$ explicitly, can be computed in the case of a symmetrically polarized soft external states up to terms that will cancel agains the remaining two integrals involving  $\mathcal{T}$ explicitly.

First we split the integral in two parts, for $l\neq j$ and $l=j$:
\ea{
I_{1,1} 
=&-\frac{\alpha'}{2} \sum_{i\neq j} (k_i\epsilon_q)(k_j\bar{\epsilon}_q)\sum_{l\neq i,j}(k_lq) \int d^2z \partial_{\bar{z}}\big[\partial_z {\cal G}(z,\,z_l) {\cal G}(z,\,z_j)\big]{\cal G}(z,\,z_i)\nonumber\\
&-\frac{\alpha'}{2} \sum_{i\neq j} (k_i\epsilon_q)(k_j\bar{\epsilon}_q)(k_jq) \int d^2z \partial_{\bar{z}}\big[\partial_z {\cal G}(z,\,z_j) {\cal G}(z,\,z_j)\big]{\cal G}(z,\,z_i)\label{1.59-2}
} 
The case $l = j$ can be rewritten as:
\ea{
I_{1,1} \big |_{l= j}
=&-\frac{\alpha'}{4} \sum_{i\neq j} (k_i\epsilon_q)(k_j\bar{\epsilon}_q)(k_jq) \int d^2z \partial_{\bar{z}}\partial_z {\cal G}^2(z,\,z_j) {\cal G}(z,\,z_i)\nonumber\\
=&
+\frac{\alpha'}{4} \sum_{i\neq j} (k_i\epsilon_q)(k_j\bar{\epsilon}_q)(k_jq) \int d^2z \partial_z {\cal G}^2(z,\,z_j) \partial_{\bar{z}}{\cal G}(z,\,z_i)\nonumber\\
=&
-\frac{\alpha'}{4} \sum_{i\neq j} (k_i\epsilon_q)(k_j\bar{\epsilon}_q)(k_jq) \int d^2z  {\cal G}^2(z,\,z_j)\partial_z \partial_{\bar{z}}{\cal G}(z,\,z_i)	\nonumber\\
=&-\frac{\alpha'}{4} \sum_{i\neq j} (k_i\epsilon_q)(k_j\bar{\epsilon}_q)(k_jq)
\left [ {2}\pi{\cal G}^2(z_i,\,z_j)+  \int {d^2z}{\cal T}(z)  {\cal G}^2(z,\,z_j) \right ]
\label{1.68-2}
}

The second case  $l \neq j$ in \Eq{1.59-2} can be computed for the soft graviton and dilaton without knowing $\G$ explicitly. Using the symmetry of the polarization tensor, we can write it as:
\ea{
I_{1,1} \big |_{l\neq j} 
=&-\frac{\alpha'}{4} \sum_{i\neq j} k_i^\mu k_j^\nu\epsilon_{q\mu\nu}^S\sum_{l\neq i,j}(k_lq) \int d^2z \partial_{\bar{z}}\big[\partial_z {\cal G}(z,\,z_l) {\cal G}(z,\,z_j)\big]{\cal G}(z,\,z_i)\nonumber\\
&-\frac{\alpha'}{4} \sum_{i\neq j} k_i^\mu k_j^\nu\epsilon_{q\mu\nu}^S\sum_{l\neq i,j}(k_lq) \int d^2z \partial_{\bar{z}}\big[\partial_z {\cal G}(z,\,z_l) {\cal G}(z,\,z_i)\big]{\cal G}(z,\,z_j)
}
and by partial integration we get:
\ea{
I_{1,1} \big |_{l\neq j}
=&
+\frac{\alpha'}{4} \sum_{i\neq j} k_i^\mu k_j^\nu\epsilon_{q\mu\nu}^S\sum_{l\neq i,j}(k_lq) \int d^2z \partial_z {\cal G}(z,\,z_l) {\cal G}(z,\,z_j)\partial_{\bar{z}}{\cal G}(z,\,z_i)\nonumber\\
&+\frac{\alpha'}{4} \sum_{i\neq j} k_i^\mu k_j^\nu\epsilon_{q\mu\nu}^S\sum_{l\neq i,j}(k_lq) \int d^2z \partial_z {\cal G}(z,\,z_l) {\cal G}(z,\,z_i)\partial_{\bar{z}}{\cal G}(z,\,z_j)\nonumber\\
=&
+\frac{\alpha'}{4} \sum_{i\neq j} k_i^\mu k_j^\nu\epsilon_{q\mu\nu}^S\sum_{l\neq i,j}(k_lq) \int d^2z \partial_z {\cal G}(z,\,z_l)\partial_{\bar{z}}\big[ {\cal G}(z,\,z_i){\cal G}(z,\,z_j)\big]
\nonumber \\
=&-\frac{\alpha'}{4} \sum_{i\neq j} k_i^\mu k_j^\nu\epsilon_{q\mu\nu}^S\sum_{l\neq i,j}(k_lq) \int d^2z\partial_{\bar{z}} \partial_z {\cal G}(z,\,z_l) {\cal G}(z,\,z_j){\cal G}(z,\,z_i)
}
Finally, we can make use of \Eq{boxG} and get
\ea{
I_{1,1} \big |_{l\neq j}=&-\frac{{2}\pi \alpha'}{4} \sum_{i\neq j} k_i^\mu k_j^\nu\epsilon_{q\mu\nu}^S\sum_{l\neq i,j}(k_lq){\cal G}(z_l,\,z_j){\cal G}(z_l,\,z_i)\nonumber\\
&-\frac{\alpha'}{4} \sum_{i\neq j} k_i^\mu k_j^\nu\epsilon_{q\mu\nu}^S(-k_iq- k_jq)\int {d^2 z}{\cal T}(z){\cal G}(z,\,z_j){\cal G}(z,\,z_i) \label{1.67-2}
}
In the last line we used $\sum_{l\neq i,j} k_lq=-(k_iq+k_jq)$.
Summing this expression with the contribution from Eq.~(\ref{1.68-2}) for a symmetrically polarized soft state we get:
\ea{
I_{1,1}
=&
-\frac{\a'}{4} \epsilon_{q\mu\nu}^S \sum_{i\neq j} k_i^\mu k_j^\nu \Bigg [
{2}\pi \sum_{l\neq i,j}(k_lq){\cal G}(z_l,\,z_j){\cal G}(z_l,\,z_i)
+{2}\pi (k_jq)  {\cal G}^2(z_i,\,z_j)
\nonumber\\
&-{\color{orange}{2(k_iq)}\int {d^2 z}{\cal T}(z){\cal G}(z,\,z_j){\cal G}(z,\,z_i)} \Bigg ]
{\color{purple}+\frac{\a'}{4} \epsilon_{q\mu\nu}^S \sum_{ j=1} k_j^\mu k_j^\nu {(k_jq)} \int {d^2z}{\cal T}(z){\cal G}^2(z,\,z_j)}
\label{1.70-2}
}
where in the last term we used $\sum_{i\neq j} ( k_i\e_q) = - ( k_j\e_q)$.
It is evident that this term exactly cancels the similar ({\color{purple} purple}) term in \Eq{1.65-2},
while the third ({\color{orange} orange}) term above for a symmetrically polarized soft state exactly cancels the similar ({\color{orange} also orange}) term in \Eq{1.65-2}.
The final result for $I_1$ when the soft state is symmetrically polarized is thus
\ea{
I_1 =&
{2}\pi \epsilon_{q\mu\nu}^S \sum_{i\neq j} \Bigg [
 \frac{k_i^\mu k_i^\nu}{k_iq} (k_jq)  {\cal G}(z_j,\,z_i)
+\frac{\alpha'}{4} k_i^\mu k_i^\nu (k_j q) {\cal G}^2(z_j,\,z_i)\nonumber\\
& +
\frac{\alpha'}{2} k_i^\mu k_j^\nu \sum_{l\neq i, j} (k_lq) {\cal G}(z_l,\,z_j)\left( {\cal G}(z_l,\,z_i)- {\cal G}(z_j,\,z_i)\right)\nonumber\\
&-\frac{\a'}{4}k_i^\mu k_j^\nu
\sum_{l\neq i,j}(k_lq){\cal G}(z_l,\,z_j){\cal G}(z_l,\,z_i)
-\frac{\a'}{4}k_i^\mu k_j^\nu (k_jq)  {\cal G}^2(z_i,\,z_j)
\Bigg]
\label{I1}
}
It is useful to rewrite the terms with sum over the label $l$ as follows:
\ea{
& \frac{{2}\pi \alpha'}{2} \epsilon_{q\mu\nu}^S \sum_{i\neq j\neq l} k_i^\mu k_j^\nu (k_lq) \Bigg [  {\cal G}(z_l,\,z_j)\left( {\cal G}(z_l,\,z_i)- {\cal G}(z_j,\,z_i)\right)
-\frac{1}{2}{\cal G}(z_l,\,z_j){\cal G}(z_l,\,z_i)
\Bigg ] \nonumber \\
&=
 \frac{{2}\pi \alpha'}{4}  \epsilon_{q\mu\nu}^S \sum_{i\neq j \neq l} 
 \left [k_j^\mu k_l^\nu (k_i q) -  k_i^\mu k_l^\nu (k_jq)- k_i^\mu k_j^\nu (k_l q)  \right ]{\cal G}(z_i,\,z_l) {\cal G}(z_i,\,z_j)
 }
The first term can be joined with the second term in \Eq{I1} by extending the sum to include $l=j$, while the third term can be joined with the last term in \Eq{I1} by extending it to include $l=j$. Thus we can write:
\ea{
I_1 =&
{2}\pi \epsilon_{q\mu\nu}^S \sum_{i\neq j} \frac{k_i^\mu k_i^\nu}{k_iq} (k_jq)  {\cal G}(z_j,\,z_i)
+ \frac{{2}\pi \alpha'}{4}  \epsilon_{q\mu\nu}^S \sum_{i\neq j } k_i^\mu k_j^\nu (k_jq) {\cal G}^2(z_i,\,z_j)
\nonumber\\
&
 +\frac{{2}\pi \alpha'}{4}  \epsilon_{q\mu\nu}^S \sum_{i\neq j , l} 
 \left [k_j^\mu k_l^\nu (k_i q) -  k_i^\mu k_l^\nu (k_jq)- k_i^\mu k_j^\nu (k_l q)  \right ]{\cal G}(z_i,\,z_l) {\cal G}(z_i,\,z_j)
\label{I1-2}
}
The result of this integral is exactly the same as the tree-level result with the difference all being in the Green function.
In particular, all dependence on ${\cal T}(z)$ have either cancelled or vanished.

\subsection{Calculation of $I_2$}
\label{I2}

The last integral to evaluate in \Eq{S1} is:
\ea{
I_2 =&\frac{\alpha'}{2} \sum_{i,j}(k_i\epsilon_q)(k_j\bar{\epsilon}_q) \sum_{l,k\neq i} (k_lq)(k_kq)  
\int d^2 z {\cal G}(z,\,z_l)\,{\cal G}(z,\,z_k) \partial_z{\cal G}( z,\,z_i) \partial_{\bar{z}} {\cal G}( z,\,z_j) e^{\frac{\alpha'}{2} k_iq {\cal G}(z,\,z_i)}
}
Since it has two factors of $q$ in front of the integral, we only need to extract the leading $1/q$ behavior of the integral.
Considering only the integral, denoted by $\mathcal{I}_{ijkl}$, let us first consider the case when $j\neq i$.
In that case we can write the integrand as
 \ea{
\mathcal{I}_{ijkl} \big |_{j\neq i}&=
\frac{\alpha'}{2}  \int d^2 z \,{\cal G}(z,\,z_l)\,{\cal G}(z,\,z_k) \partial_z{\cal G}( z,\,z_i) \partial_{\bar{z}} {\cal G}( z,\,z_j) e^{\frac{\alpha'}{2} k_iq {\cal G}(z,\,z_i)}\nonumber\\
&=\frac{1}{ k_iq}\int d^2 z \,{\cal G}(z,\,z_l)\,{\cal G}(z,\,z_k) \partial_{\bar{z}} {\cal G}( z,\,z_j)\partial_z e^{\frac{\alpha'}{2} k_iq {\cal G}(z,\,z_i)}\nonumber\\
&=
-\frac{1}{ k_iq}\int d^2 z \,\partial_z\Big[{\cal G}(z,\,z_l)\,{\cal G}(z,\,z_k) \partial_{\bar{z}} {\cal G}( z,\,z_j)\Big] e^{\frac{\alpha'}{2} k_iq {\cal G}(z,\,z_i)}
\nonumber \\
&= 0 + \Ord(q^0)
}
where in the last equality we expanded the exponential (since in the bracket there is no singularity at $z=z_i$), and the leading term vanishes because it is a total derivative.

To compute the case $i= j$ observe that
\ea{
&\partial_z\partial_{\bar{z}} e^{\frac{\alpha'}{2} qk_i {\cal G}(z,\,z_i)}=\partial_z\Big[ \frac{\alpha'}{2} k_iq\, \partial_{\bar{z}}{\cal G}(z,\,z_i)\, e^{\frac{\alpha'}{2} qk_i{\cal G}(z,\,z_i)}\Big]
\nonumber\\
&=\frac{\alpha'}{2} k_iq\,\partial_z\partial_{\bar{z}} {\cal G}(z,\,z_i) e^{\frac{\alpha'}{2} k_iq {\cal G}(z,\,z_i)}+\left(\frac{\alpha'}{2}k_iq \right)^2 \partial_{\bar{z}}{\cal G}(z,\,z_i) \partial_z{\cal G}(z,\,z_i)\,  e^{\frac{\alpha'}{2} k_iq {\cal G}(z,\,z_i)}
}
Thus we can make the following rewriting of the integral:
\ea{
\mathcal{I}_{ijkl} \big |_{j=i}
=& \frac{1}{\frac{\alpha'}{2} \left(qk_i\right)^2}\int d^2 z \,{\cal G}(z,\,z_l)\,{\cal G}(z,\,z_k)\partial_z\partial_{\bar{z}}e^{\frac{\alpha'}{2} qk_i{\cal G}(z,\,z_i)}\nonumber\\
&-\frac{1}{qk_i} \int d^2z\, {\cal G}(z,\,z_l)\,{\cal G}(z,\,z_k)
\partial_z\partial_{\bar{z}}{\cal G}(z,\,z_i) e^{\frac{\alpha'}{2} qk_i{\cal G}(z,\,z_i)}\nonumber\\
=
&- \frac{1}{\frac{\alpha'}{2}\left(qk_i\right)^2}\int d^2 z \,\partial_z\Big[{\cal G}(z,\,z_l)\,{\cal G}(z,\,z_k)\Big]\partial_{\bar{z}}e^{\frac{\alpha'}{2} qk_i{\cal G}(z,\,z_i)}\nonumber\\
&
-\frac{1}{qk_i}\int {d^2z}{\cal T}(z)\, {\cal G}(z,\,z_l)\,{\cal G}(z,\,z_k)
e^{\frac{\alpha'}{2} qk_i{\cal G}(z,\,z_i)}
\label{1.72-2}
}
where total derivatives were set to zero and we made use of the identity \eqref{eGzero}.
We can now expand the exponentials, since $i \neq l, k$, and the leading order contributions read:  
\ea{
\mathcal{I}_{ijkl} \big |_{j=i}= &\frac{1}{\frac{\alpha'}{2}qk_i}\int d^2 z \,\partial_{\bar{z}}\partial_z\Big({\cal G}(z,\,z_l)\,{\cal G}(z,\,z_k)\Big){\cal G}(z,\,z_i)
\nonumber \\
&
-\frac{1}{\frac{\alpha'}{2} qk_i} \int {d^2z}{\cal T}(z)\, {\cal G}(z,\,z_l)\,{\cal G}(z,\,z_k)+O(q^0)
}
A double integration by parts of the first terms gives:
\ea{
\mathcal{I}_{ijkl} \big |_{j=i}= &\frac{{2}\pi}{ \frac{\alpha'}{2} qk_i}{\cal G}(z_i,\,z_l)\,{\cal G}(z_i,\,z_k){+\frac{1}{\frac{\alpha'}{2}qk_i}\int {d^2 z}{\cal T}(z) \,{\cal G}(z,\,z_l)\,{\cal G}(z,\,z_k)}\nonumber\\
&{-\frac{1}{\frac{\alpha'}{2} qk_i}\int {d^2 z}{\cal T}(z)\, {\cal G}(z,\,z_l)\,{\cal G}(z,\,z_k)}+O(q^0)
}
Hence the two terms involving ${\cal T}$ cancel.

We have thus extracted all $1/q$ dependence of the integral and the final result for $I_2$ reads:
\ea{
I_2 
&={2}\pi \sum_{i=1}(k_i\epsilon_q)(k_i\bar{\epsilon}_q) \sum_{l,k\neq i} \frac{ (k_lq)(k_kq)    }{qk_i}{\cal G}(z_i,\,z_l)\,{\cal G}(z_i,\,z_k)
}
This result is formally identical with the tree-level result with all the difference being in the Green function.

\section{Calculation of $S_2^{(1)}$}
\label{X}
In this appendix we consider, at one loop, the integral that appears in Eq.~(\ref{S21}). 
For the calculation we have made explicit use of the one loop expression, and we have not been able to extend this to the generic multiloop level.

The integral in Eq.~(\ref{S21}) can be easily evaluated  by performing the change of variable  $z=e^{2\pi i\nu}$, and by using the one-loop expression of Green's function given for example in Eq.~(3.8) of Ref.~\cite{9910056}. The result of the calculation is an expression depending only on the moduli of the torus that makes Eq.~(\ref{S21}) zero on-shell,  since $\sum_{l=1}^N k_lq=-q^2=0$. 
	In the following instead of giving the details of the calculation,  we will prove the independence of the integral on the variable $z_l$, i.e.: 
	\begin{eqnarray}
		\partial_{z_l}\int_{\cal A} d^2z \omega(z)\bar{\omega}(\bar{z})
		{\cal G}_1(z,\,z_l)=0\,,\label{1loopsubsuba}
	\end{eqnarray}
	This is sufficient to show the vanishing of Eq.~(\ref{S21}) on shell.
	Here ${\cal A}=\{ z\in \mathbb{C},\, \mbox{s.t.}\, |\mk|\leq|z|\leq 1\}$ denotes the one-loop integration region.  
	In order to prove Eq.~(\ref{1loopsubsuba}), we first 
	observe that, with the choice of $V'_i(0)=2\pi z_i$, the Green function satisfies the identity~\cite{DiVecchia:NP96}:
	\begin{eqnarray}
		\omega(z_l)\partial_{z} {\cal G}_1(z,\,z_l)+\omega(z)\partial_{z_l} {\cal G}_1(z,\,z_l)=0\label{id1loop}
	\end{eqnarray}
	that, when used in Eq.~(\ref{1loopsubsuba}), gives:
	\begin{eqnarray}
		-\int_{\cal A} d^2z
		\partial_z\frac{{\cal G}_1(z,\,z_l)}{\bar{z} z_l} =-\frac{1}{2iz_l} \Big[\oint_{|z|=1}\frac{d\bar{z}}{ \bar{z}}{\cal G}_1(z,\,z_l)-\oint_{|z|=|\mk|}\frac{d\bar{z}}{\bar{z}} {\cal G}_1(z,\,z_l)\Big]
	\end{eqnarray} 
In the right-hand-side of the previous expression  one can change the variable $\bar{z}=\bar{z}'\, \bar{\mk}$, getting:
	\begin{eqnarray}
		-\int_{\cal A} d^2z
		\partial_z\frac{{\cal G}_1(z,\,z_l)}{\bar{z} z_l}=-\frac{1}{2iz_l} \oint_{|z|=1}\frac{d\bar{z}}{\bar{z}} [{\cal G}_1(z,\,z_l)-{\cal G}_1(kz,\,z_l)]=0
	\end{eqnarray}
	The expression is zero from the invariance of the Green function under transport  around the $b$-cycle of the torus, i.e. ${\cal G}_1(z,\,z_l)={\cal G}_1(kz,\,z_l)$.}

The multiloop  extension of such a proof  would require the multiloop generalization  of  Eq.~(\ref{1loopsubsuba}). At one loop this follows from the identity (\ref{id1loop}) and from the invariance of the Green function along the homology cycles of the Riemann surface. The main obstacle to extend these considerations to arbitrary orders in the perturbative expansion, is the lack of a multiloop identity similar to Eq.~(\ref{id1loop}).

\end{document}